\documentclass[sigconf]{acmart}
\AtBeginDocument{%
  }

\copyrightyear{2026}
\acmYear{2026}
\setcopyright{cc}
\setcctype{by}
\acmConference[CHI '26]{Proceedings of the 2026 CHI Conference on Human Factors in Computing Systems}{April 13--17, 2026}{Barcelona, Spain}
\acmBooktitle{Proceedings of the 2026 CHI Conference on Human Factors in Computing Systems (CHI '26), April 13--17, 2026, Barcelona, Spain}
\acmPrice{}
\acmDOI{10.1145/3772318.3790738}
\acmISBN{979-8-4007-2278-3/2026/04}

\begin{document}

\title[Towards Inclusive External Human-Machine Interface]{Towards Inclusive External Human-Machine Interface: Exploring the Effects of Visual and Auditory eHMI for Deaf and Hard-of-Hearing People}

\author{Wenge Xu}
\orcid{0000-0001-7227-7437}
\email{Wenge.Xu@bcu.ac.uk}
\affiliation{
 \institution{Birmingham City University}
 \city{Birmingham}
 \country{United Kingdom}
}

\author{Foroogh Hajiseyedjavadi}
\orcid{0000-0003-3448-1239}
\email{Foroogh.Hajiseyedjavadi@bcu.ac.uk}
\affiliation{%
 \institution{Birmingham City University}
 \city{Birmingham}
 \country{United Kingdom}
}

\author{Kurtis Weir}
\orcid{0000-0002-7094-4600}
\email{Kurtis.Weir@bcu.ac.uk}
\affiliation{
 \institution{Birmingham City University}
 \city{Birmingham}
 \country{United Kingdom}
}

\author{Chukwuemeka Eze}
\email{Chukwuemeka.Eze@mail.bcu.ac.uk}
\affiliation{
 \institution{Birmingham City University}
 \city{Birmingham}
 \country{United Kingdom}
}

\author{Mark Colley}
\orcid{0000-0001-5207-5029}
\email{m.colley@ucl.ac.uk}
\affiliation{%
 \institution{UCL Interaction Centre}
 \city{London}
 \country{United Kingdom}
}

\renewcommand{\shortauthors}{Xu et al.}

\begin{abstract}
External Human-Machine Interfaces (eHMIs) have been proposed to facilitate communication between Automated Vehicles (AVs) and pedestrians. However, no attention was given to Deaf and Hard-of-Hearing (DHH) people. We conducted a formative study through focus groups with 6 DHH people and 6 key stakeholders (including researchers, assistive technologists, and automotive interface designers) to compare proposed eHMIs and extract key design requirements. Subsequently, we investigated the effects of visual and auditory eHMI in a virtual reality user study with 32 participants (16 DHH). Results from our scenario suggesting that (1) DHH participants spent more time looking at the AV; (2) both visual and auditory eHMIs enhanced trust, usefulness, and perceived safety; and (3) only visual eHMIs reduced the time to step into the road, time looking at the AV, gaze time, and percentage looking at active visual eHMI components. Lastly, we provided five practical implications for making eHMI inclusive of DHH people.
\end{abstract}




\maketitle

\section{Introduction} 
External human-machine interfaces (eHMI) have been proposed for several years to address the absence of traditional explicit communication cues between pedestrians and human drivers~\citep{SUCHA201741} and assist in communication between pedestrians and automated vehicles (AVs)~\citep{smartphone}. However, a significant gap in the eHMI literature is that most papers only focus on non-disabled people, excluding 1.3 billion (16\%) of the world population who live with significant disability\footnote{\href{https://www.who.int/news-room/fact-sheets/detail/disability-and-health}{WHO: Disability and Health}; accessed 14.04.2025}. The disability of pedestrians is a critical contributory factor to fatal or serious collisions with vehicles\footnote{\href{https://www.gov.uk/government/statistics/reported-road-casualties-great-britain-pedestrian-factsheet-2022}{Reported road casualties in Great Britain: pedestrian factsheet, 2022}; accessed 14.08.2024}. The design of eHMIs needs to consider the needs of disabled people; otherwise, it may be inaccessible to them, creating a further divide and inequality in transport. There are initial explorations towards making eHMIs accessible for low vision or blind people~\citep{Mark_Vision}, intellectually disabled people~\citep{10.1145/3546717}, and wheelchair users~\citep{Asha_wheelchair}. However, there is a lack of investigation on deaf and hard-of-hearing (DHH) people. According to the World Health Organisation, the population of DHH people reached 430 million worldwide and is estimated to reach 700 million in 2050 (i.e., 1 in every 10 people)\footnote{\href{https://www.who.int/news-room/fact-sheets/detail/deafness-and-hearing-loss}{WHO: Deafness and Hearing Loss}; accessed 14.04.2025}. 

Most existing work on eHMIs focused on visual communication design patterns (e.g., abstract light~\citep{Bumper_Light}, anthropomorphic features~\citep{Eyes}, text~\citep{text_icons}, symbols~\citep{symbol}, or projection~\citep{Chris_project}). Providing only visual eHMIs would not benefit low vision or blind people or those who experience situational impairments (e.g., being distracted by secondary activities, occluded view)~\citep{10.1145/3349263.3351523}. Several works have suggested that using multi-modal eHMI, e.g., combined visual-auditory communication, could make AVs accessible and benefit more pedestrians~\citep{Mark_Vision,10.1145/3546717}. However, what visual designs would work for DHH people remains unclear due to the lack of involvement of DHH in the design and evaluation phase. Additionally, how DHH people would perceive the use of auditory communication, such as speech, remains unclear. 

We first conducted a focus group study with DHH people and key stakeholders (i.e., human factor researchers, eHMI researchers, accessibility researchers, assistive technologist, and HMI designer) to help us (1) minimise the visual design candidates to be used in our VR study, and (2) come up with initial design requirements. We then conducted a mixed-design VR study with N=32 participants to investigate the effect of Visual (\textit{No Visual}, \textit{Abstract Light}, \textit{Abstract Light + Text}, \textit{Abstract Light + Symbol}) and Auditory (\textit{Without Speech} and \textit{With Speech}) eHMIs among Hearing group (N=16) and DHH group (N=16) with regards of their crossing experience (trust, acceptance, perceived safety, mental load) and behaviour (gaze, step into the road time, early step into the road count). Importantly, we further investigated the extent to which the DHH group valued these differently from the Hearing group.

\textit{Contribution Statement:} Our main contributions: (1) A first focus group study with DHH people (N=6) and relevant stakeholders (N=6) comparing six visual external communication designs from the literature and exploring initial design requirements. (2) A first VR user study that involves DHH participants (16 out of 32 participants) showing that (i) there are behaviour (i.e., eye gaze) differences between participants in DHH group and Hearing group, (ii) providing visual communication improves crossing experience (i.e., trust, usefulness, perceived safety) and behaviour (step-in road time, time spent on the AV, time and percentage of gaze spent on active visual eHMI components), and (iii) providing speech as auditory communication improves crossing experience (i.e., trust, usefulness, perceived safety). (3) 5 practical implications that pave the way for future eHMI design and research.

\section{Related Work}
\subsection{DHH People and Road Crossing}
Hearing and eyesight are the most relevant senses to act adequately in traffic situations~\citep{deafblind}. Challenges such as obstructed views due to parked vehicles, low lighting, adverse weather conditions, or dazzling produced by adjacent lights could affect the crossing visually~\citep{wearther_lighting, crossing_visual}. Hearing helps us understand the direction and distance of sounds, allowing us to judge the location of potential threats or obstacles~\citep{kolarik2015auditory}, which could be particularly challenging for DHH people~\citep{hearing_aids_can_fail}. Previous study with DHH teenagers showed that almost half of DHH teenagers have been involved in traffic accidents, which is a much higher rate than hearing teenagers \citet{GUR2021100994}.

Hearing technologies such as hearing aids and hearing implants are the most commonly used technologies by DHH people \citep{WHO2021HearingReport}. Hearing aids are intended to help people with mild to moderate hearing loss\footnote{\url{https://www.nidcd.nih.gov/health/hearing-aids}; accessed 02.08.2024}. Hearing aids may improve sound localisation, but have failed to improve it consistently and could even impair it~\citep{hearing_aids_can_fail}. As the perceived benefits not meeting the expectation and the wearing discomfort, only about 20\% people who would benefit from hearing aids use one~\citep{McCormack01052013}. Hearing implants such as cochlear implants help provide a sense of sound to people with severe to profound deafness\footnote{\url{https://www.nidcd.nih.gov/health/cochlear-implants}}. This technology does not restore normal hearing; instead, they restore some sensation of auditory perception to help deaf people understand sounds or speech~\citep{Cochlear_implants}. However, data from the UK showed that only 1.3\% of severely or profoundly deaf people use hearing implants \footnote{Hearing loss statistics in the UK: 
\url{https://www.hearinglink.org/your-hearing/about-hearing/facts-about-deafness-hearing-loss/}}. In short, neither technology is perfect, nor would it help DHH people address the absence of hearing cues, which may lead to missing critical information when interacting with AVs through eHMIs, especially if they employ auditory feedback.

Limited work has been conducted to understand road crossing among DHH pedestrians. \citet{street_crossing_disability} conducted a field study in an urban environment with manually driven vehicles and found that DHH people often experienced heightened apprehension when initiating a crossing and exercised greater caution toward approaching vehicles to ensure safety. \citet{hearing_aid_location} suggest that pedestrians with moderate deafness are at a higher risk of being injured by a vehicle because they have difficulty in identifying the direction the sound is coming from. Overall, the lack of access to auditory information has reduced feelings of safety among DHH people~\citep{deaf_road_safety}. Over time, this could constitute a negative and fatiguing experience, which may discourage active travel like walking~\citep{Active_travel}, resulting in reduced physical activity levels~\citep{Carlin2016} and broadening inequities in mobility~\citep{inequality_mobility_walking}. 

Regarding crossing strategies, \citet{street_crossing_disability} observed that DHH pedestrians employed deliberate tactics to communicate with drivers to secure their right-of-way. These included actively seeking direct visual contact with approaching drivers and using hand gestures to capture attention and signal them to slow down or stop. Such cooperative, bidirectional interaction allowed DHH pedestrians to create safer crossing conditions. However, this strategy poses a critical challenge in AVs due to the absence of human drivers. This highlights the need for investigation visual and auditory eHMIs with DHH people.

\subsection{External Human-Machine Interface}
As eyesight (visual) and hearing (auditory) are the most relevant senses (modalities) to act adequately~\citep{deafblind}, we first discuss visual eHMIs and auditory eHMIs, then move on to discussing eHMIs and disabilities, and finish with multi-modal eHMIs as a potential solution for disabled people when interacting with AVs.

\subsubsection{Visual eHMIs}
\citet{DEY_Taming} found that the visual modality is the most commonly adopted modality among the eHMI literature, and multiple studies \cite{doi:10.1177/0018720819836343,10.1007/978-3-319-41682-3_41} have shown that the presence of a visual eHMI could assist in the communication between vulnerable road users and AVs. These proposed visual eHMI concepts can be categorized into the following six visual design patterns, including (1) Road-based Projection~\citep{Chris_project}, (2) Symbols~\citep{symbol}, (3) Text~\citep{text_icons}, (4) Anthropomorphic, including Smileys\footnote{\href{https://www.semcon.com/article/the-smiling-car}{Semcon: The Smiling car}; accessed 14.07.2024}, eyes~\citep{Eyes}, and gestures~\citep{smartphone}, (5) Abstract Light: one-dimensional light segment~\citep{Bumper_Light} or two-dimensional light display, and (6) Situational Awareness: a light tracker to show the situational awareness of the AV~\citep{DEY_Taming}. 

\citet{Dey_scalable} compared Situational Awareness, Progress Bar (Abstract Light), Moving Light Bar (Abstract Light), and Road-based Projection and found that Road-based Projection is preferred when the AV has to interact with multiple pedestrians. Text was found preferred by participants in a video-based study \citet{ACKERMANN2019272} and a survey study~\citep{text_icons}. Despite much research into comparing visual concepts, there is still no explicit agreement on which visual design is the best and how many visual signals are sufficient~\cite{multi_modal_dey}. Due to the lack of agreement towards the choice of visual design and the gap in the research with DHH people, we first conducted a formative study to understand DHH people and the needs of relevant stakeholders and to narrow down visual eHMI designs that we will be using in the VR study. Nevertheless, we locate the visual eHMI on the front of the vehicle grill, which mimics the current expectations and living experience of pedestrians, who generally look towards the location of the driver’s head or vehicle movement~\citep{info11010013,10.1145/3342197.3344523}, but also follows the prior works~\citep{multi_modal_dey,scalability_Mark}. 

\subsubsection{Auditory eHMI}
There is relatively less attention to the auditory eHMI design for AVs. \citet{DEB2018135} suggested that visual eHMIs had a much larger effect on the willingness to initiate crossing when compared to auditory eHMIs (horn, music, and verbal warning saying "safe to cross"). However, auditory eHMI holds several benefits, such as announcing situation awareness and detection~\cite{MERAT2018244} and is a reliable source for vision disabled people~\cite{Mark_Vision}. The implementations of auditory eHMI are either through speech (verbal messages) \citet{Mahadevan_audio, DEB2018135, Hudson_verbal} or non-speech auditory signals (e.g., jingles, humming, bells)~\citep{Bell_sound,multi_modal_dey, Florentine}. 

\citet{multi_modal_dey} compared two non-speech concepts (i.e., bell and droning) in a video-based study, and found that participants had completely different associations and mental models for using these two sounds. Many people perceived the bell as a calm, inviting, and friendly signal, but others perceived it as urgent and rushed and associated it with a warning. Similarly, while several people felt the choice of drone was an intuitive and natural concept for the vehicle’s engine sound and speed, others found it unpleasant and burdensome. Prior works have suggested that verbal messages might be more reliable than non-speech auditory signals. \citet{DEB2018135} investigated multiple sound-related features in a VR study, finding that the verbal message was the most favoured audible feature compared to horn, music, and no sound. \citet{Hudson_verbal} found similar results where verbal messages are preferred over music. In addition, verbal messages may be more inclusive for vision disabled people~\citep{Mark_Vision}. Therefore, we used speech in our VR study to explore DHH people's opinions regarding using auditory eHMI. 

\subsubsection{eHMI and Disabilities}
Despite extensive research having been conducted on the development of eHMIs, only limited research has been conducted with disabled people. Some works focus on co-designing with disabled people from scratch, whereas others focus on evaluating existing concepts and tailoring them for disabled people. \citet{Asha_wheelchair} focused on the design of eHMI tailored for wheelchair users; they employed inclusive design practices to work with a wheelchair user to explore the user requirements of wheelchair users when communicating with AVs as pedestrians. On the other hand, \citet{Mark_Vision} evaluated existing auditory concepts with the low vision and blind people through a workshop study and found that the speech auditory eHMI was best received. Their follow-up VR study suggests that having more content in the speech message could reduce the mental load. \citet{10.1145/3546717} compared eHMI concepts (baseline, visual-only, auditory-only, multi-modal---both visual and audio) through an online video-based survey study with intellectual disabled participants and non-intellectual disabled participants; they found that auditory eHMIs performed worse than visual or multi-modal eHMIs. In general, \citet{10.1145/3546717} found that multi-modal eHMIs positively affect quality and inclusion. 

The need for multi-modal eHMIs has been a common agreement among the studies worked with disabled people~\citep{Asha_wheelchair,mark_include_impairment, Mark_Vision, 10.1145/3546717}. In addition, it has been suggested by a review paper on eHMI accessibility~\citep{10555703}, a core reason is that each modality has specific trade-off \citet{smartphone}. Multi-modal interfaces or feedback designs have been commonly used by other fields to make the interaction accessible for disabled people~\citep{Argyropoulos2008,10.1145/638249.638260, Covarrubias01072014}. 

\subsubsection{Multi-Modal eHMIs}
The literature suggests that multi-modal eHMI brings several benefits to the pedestrian. \citet{sym13040687} conducted a VR study to evaluate 12 eHMI concepts, in combinations of visual (smile/arrow), audio (human voice/warning sound), and vehicle movement style (the approaching speed decreases gradually/remains unchanged), and concluded that multi-modal eHMIs resulted in more satisfactory interaction and improved safety compared to the unimodal eHMI. 
\citet{He_VR} conducted a VR study and found that audio-visual modality (symbol and anthropomorphic voice) was more appealing than the eHMI with a unimodal modality (visual or auditory). Results from the Wizard-of-Oz study by~\citep{10.1007/978-3-030-78645-8_27} suggested that a combination of audio-visual modality is most effective in understanding information. 

Despite these potential benefits discussed in the literature, a recent video-based study by \citet{multi_modal_dey} found that multi-modal eHMI do not outperform unimodal eHMI regarding effectiveness and user experience. Their qualitative data suggested that multi-modal eHMI may also overwhelm pedestrians, although it also offers assurance of vehicle intention. To our knowledge, no studies have been conducted on exploring eHMIs with DHH people. Therefore, we conducted a VR study to explore the visual eHMI designs, whether auditory eHMIs would be well perceived by DHH people, and how DHH people may perceive eHMIs differently from hearing people.

\section{Formative Study}
As there is a lack of work in eHMI with Deaf and Hard-of-Hearing (HoH) people, we first conducted a formative study with two focus groups (a Deaf-focused group for profoundly deaf people and a HoH-focused group for mild to severe hearing loss people) to gain insights from DHH people and key stakeholders (1) regarding high-level visual eHMI design requirements and (2) about visual eHMI candidates for our VR-based user study. 

Focus groups are well-suited to the exploratory nature of formative study and are commonly used in early-stage HCI studies~\citep{10.1145/3706598.3713542,10.1145/3491102.3502077,10.1145/3706598.3713409}. We intentionally brought together DHH people and key stakeholders to ensure a wide range of perspectives and to facilitate more inclusive discussions about current accessibility challenges in pedestrian–AV interaction~\citep{Bouma18122022,Creed17102024,Duarte}. The group format enabled us to capture a breadth of viewpoints and to observe how DHH people and stakeholders collectively interpret, negotiate, and debate these concepts.

\subsection{Participants}
Twelve participants (4F, 8M, mean age=35.17, SD=9.20) participated in the study (see \autoref{table:participants}). The DHH people were recruited through posters, social media platforms, and word-of-mouth. Stakeholders like eHMI researchers, accessibility researchers, HMI industry specialists, and people who work in the charity sector were recruited through research papers, charity networks, social media platforms, and prior project databases. The group allocation was done based on participants' lived experience, (2) research interest, or (3) industry expertise. The hearing loss levels were categorised based on recommendations from National Health Services\footnote{National Health Service: \url{https://www.esht.nhs.uk/service/audiology/diagnosis-and-testing/}}: normal (<20 dB), mild (21 - 40 dB), moderate (41 - 70 dB), severe (71 - 95 dB), and profound (>95 dB).

\begin{table}[ht]
\caption{Pseudonyms and background of focus group participants}
\begin{tabular}{p{0.15\linewidth} p{0.13\linewidth} p{0.6\linewidth}}
\hline
    Pseudonym & Age & Relatedness to the project\\
    \hline
    D1 & 30s & Profoundly deaf in both ears and uses BSL \\
    D2 & mid-20s & Profoundly deaf in both ears and uses BSL\\
    D3 & 50s & Human factors and eHMI researcher who has worked with disabled people \\
    D4 & 30s & Human factors and eHMI researcher who works at a national centre for accessible transport \\
    D5 & 40s & HMI designer who works at JLR and has a special interest in eHMI and AV\\
    D6 & 30s & Accessibility researcher who has single-ear profound deafness and has research interests in DHH people\\
    \hline 
    H7 & 20s & Severe deafness in both ears \\
    H8 & late-30s & Mild deafness in a single ear\\
    H9 & 30s & Severe deafness in one ear and moderate deafness in the other, who is also an assistive technologist \\
    H10 & late-40s & Human factor and eHMI researcher who has worked with JLR \\
    H11 & mid-30s & Accessibility researcher with a special interest in HoH research \\
    H12 & mid-30s & Accessibility researcher with a special interest in HoH research who works for a national charity for DHH people\\
    \hline
    \end{tabular}
    \footnotesize Pseudonyms consist of focus group initial (D: deaf, H: Hard-of-Hearing) and Participant ID (1-12).

  \label{table:participants}
  \Description{Table shows demographic of participants participated in the focus group study, with columns from left to right shows, their allocated ID, age, relatedness to the project. }
\end{table}

\subsection{Apparatus}
Similar to prior focus group studies~\citep{Creed17102024, 10.1145/3609230, 10.1145/3653689}, we used Teams as the online video conferencing tool. We ensured the accessibility of Teams by asking participants about any accessibility adjustments prior to the study. Transcripts were enabled by default for all users. Two British Sign Language (BSL) interpreters were recruited for BSL users in the Deaf group; interpreters were pinned to the screen for BSL users. Overall, participants found the choice of Teams acceptable. 

\subsection{Evaluated Concepts}
We evaluated the following six commonly used eHMI visual concepts~\citep{multi_modal_dey}; each concept comes with a video demonstration. 

\begin{enumerate}
\item \textit{Abstract Light}: This concept employs a light bar on the bumper to communicate the AV's state and intention~\citep{Bumper_Light}. The video used in the study had the following implementation: While the AV is driving, the light bar is glowing statically in cyan. When yielding for a pedestrian, the bar starts to pulsate between on and off at 1 Hz. Once the AV comes to a complete stop, the light bar turns off.

\item \textit{Situational Awareness}: This concept follows the design pattern from ~\citep{DEY_Taming}, where a square cyan indicator at the bottom of the windscreen of the AV would appear when it detects the pedestrian and would point toward the pedestrian to show the AV intends to yield for~\citep{scalability_Mark}. If the pedestrian steps onto the road and starts crossing, the cyan indicator will follow the pedestrian. 

\item \textit{Text}: The \textit{Text} concept displays text to the pedestrian on a bumper display~\citep{scalability_Mark}. While driving, the bumper display will show "DRIVING". When an AV begins to yield, the text will switch to "YIELDING". When stopped, it displays "STOPPED".

\item \textit{Symbol}: This concept follows the symbol design pattern~\citep{DEY_Taming, symbol} displayed on a bumper display. The display does not show any additional information and would activate a pedestrian symbol to indicate the AV would yield for the pedestrian~\citep{scalability_Mark}.

\item \textit{Road-based Projection}: This concept was based on the prior works in projection design pattern~\citep{Projection3, Projection2}. Once the AV begins to yield, it projects a rectangle with triangles representing arrows up to 10 m in front of the car. The projection moves with the AV until it is 10 m away from its yielding position. At this point, the end of the rectangle represents the line that marks the spot where the AV will stop. The AV will continue to come closer to the yielding line. Once stopped, the stopping line will remain present~\citep{scalability_Mark}.

\item \textit{Anthropomorphic}: We simplified the design patterns introduced by \citet{DEY_Taming} by combining "Eyes", "Smiling", and "Other Anthropomorphic" into \textit{Anthropomorphic}. For the video presentation of this eHMI design, we implement the same Smiling AV concept as described in~\citep{scalability_Mark}, where the anthropomorphic interface is displayed through the AV’s bumper display with a "mouth" on it. If the AV is not yielding, the mouth remains its natural expression (i.e., a horizontal line). If the AV is yielding to a pedestrian, the mouth turns the horizontal line into a smile.
\end{enumerate}

\subsection{Procedure}
Both focus group sessions were conducted on the same day, with each participant attending only their assigned group. The study began with an introduction from the research team, outlining the research brief and key activities. Upon transitioning to the eHMI section, each group’s facilitator (part of the author team) provided a high-level overview of eHMI (e.g., what eHMI is and why it is needed) followed by a question and answer session for participants. Then, the group discussion began with six eHMI concepts discussed in order. For each concept, the facilitator first gave a brief introduction accompanied by images and a short video presentation. Participants were then asked to take turns sharing their views on the benefits, disadvantages, concerns, and potential barriers of the concept, with the reminder that there were no right or wrong answers. Once all participants shared the corresponding eHMI idea, they can freely discuss the concept. Before moving on to the next concept, the facilitator asked if participants had any additional comments. 

This activity took around 150 minutes, including breaks. After participants went through all concepts, they filled in a questionnaire (see Section~\ref{sec:outcome_measure}). After completing the questionnaire, they had another 30 minutes for open discussion (if they had other design ideas, auditory information).

\subsection{Quantitative Measurements}\label{sec:outcome_measure}
We employed the following questions to help us decide candidates for the subsequent VR user study.

\begin{enumerate}
 \item Trust: Trust was measured through the Trust subscale and the Understandability subscale from the Trust in Automation questionnaire~\cite{trust_korber} in a 5-point Likert scale (1=Strongly disagree to 5=Strongly agree). 
 \item Acceptance: We employed the van der Laan acceptance scale with the subscales "usefulness" and "satisfying"~\cite{VANDERLAAN19971}. The scale is measured in a 5-point Likert scale during collection and adjusted to -2 (negative rating) to +2 (positive rating) in the data analysis.
 \item Perceived Safety: We assessed the Perceived Safety using questions from ~\cite{safety_Faas} with a 7-point Likert scale, from -3 (negative) to +3 (positive).
\item Mental Workload: We used the raw Mental Workload subscale on a 21-point Likert scale from the NASA-TLX Questionnaire~\citep{HART1988139}.
 \item Necessity and Reasonability~\citep{scalability_Mark}: We employed a 7-point Likert scale (1=Totally Disagree to 7=Totally Agree) to measure participants' perception of the Necessity and Reasonability of overall visual eHMI concept usage. 
 \item Rank: Participants ranked the eHMI concepts regarding their preference (the lower the number, the better). 
\end{enumerate}

\subsection{Findings}
This formative study helped us identify (1) appropriate visual eHMI candidates for the VR user study through a combination of questionnaire data and ranking, and (2) top-level design requirements through qualitative data. 

\subsubsection{Quantitative Survey Results and Discussion}
We used Shapiro-Wilk to check the normality of the data, for normally distributed data, we used one-way repeated measure ANOVA with eHMI Types as the within-subject factor to help us gain initial understanding of the DHH people and relevant stakholders' perspectives regarding each eHMI design. We used the Friedman test to analyze data if they were not normally distributed. Bonferroni corrections were always used in all post-hoc analyses. 

\textbf{Trust in Automation}. Regarding Understandability, an ANOVA revealed a significant effect of eHMI Types ($F(5,55) = 7.864, p < .001, \eta_{p}^{2}=.417$). Post-hoc pairwise comparisons indicated that \textit{Abstract Light} ($M=3.10, SD=0.77$) was rated higher than \textit{Anthropomorphic} ($M=1.96, SD=0.96, p = .029$) and \textit{Situational Awareness} ($M=1.90, SD=0.76, p=.013$). \textit{Text} ($M=3.40, SD=0.90$) was rated higher than \textit{Anthropomorphic} ($M=1.96, SD=0.96, p=.034$) and \textit{Situational Awareness} ($M=1.90, SD=0.76, p=.010$).

As for Trust, ANOVA yielded a significant effect of eHMI Types ($F(5,55) = 8.907, p < .001, \eta_{p}^{2}=.447$). Post-hoc tests suggested that participants rated a significant lower rating for \textit{Situational Awareness} ($M=1.96, SD=0.72$) than \textit{Abstract Light} ($M=3.29, SD=0.81, p=.004$), \textit{Symbol} ($M=3.30, SD=0.92, p =.031$) and \textit{Text} ($M=3.50, SD=0.74, p=.003$). \textit{Anthropomorphic} was also rated lower than \textit{Abstract Light} ($M=3.29, SD=0.81,p=.028$) and \textit{Text} ($M=3.50, SD=0.74, p=.032$).

\textbf{Perceived Safety}.
ANOVA revealed a significant effect of eHMI Types $F(5,55) = 7.305, p < .001, \eta_{p}^{2}=.399$. Post-hoc tests showed that \textit{Situational Awareness} ($M=-1.77, SD=0.97$) was rated significantly lower than \textit{Symbol} ($M=0.69, SD=1.43, p=.007$), \textit{Abstract Light} ($M=0.42, SD=1.47, p=.014$), \textit{Text} ($M=0.71,SD=1.45, p=.002$), \textit{Road-based Projection} ($M=0.02, SD=1.27, p=.022$).

\textbf{Acceptance}. As for Usefulness, ANOVA showed a significant effect of eHMI Types $F(5,55) = 7.453, p < .001, \eta_{p}^{2}=.404$. Post-hoc tests suggested that \textit{Situational Awareness} ($M=-0.85,0.73$) as rated significantly lower than \textit{Symbol} ($M=0.63, SD=0.90, p=.020$), \textit{Abstract Light} ($M=0.50, SD=0.69, p=.003$), and \textit{Text} ($M=0.80, SD=0.72, p=.001$).

Regarding Satisfying, ANOVA yielded a significant effect of eHMI Types $F(5,55) = 8.728, p < .001, \eta_{p}^{2}=.442$). Post-hoc tests suggested that \textit{Situational Awareness} ($M=-1.02, SD=0.75$) was rated significantly lower than \textit{Symbol} ($M=0.67, SD=0.83, p=.006$), \textit{Abstract Light} ($M=0.33, SD=0.69, p=.002$), and \textit{Text} ($M=0.50, SD=0.75, p=.011$).

\textbf{Mental Workload}. ANOVA yielded a significant main effect of eHMI Types ($F(5,55) = 7.105, p < .001, \eta_{p}^{2}=.392$) on Mental Workload. Post-hoc pairwise comparisons tests suggested that \textit{Situational Awareness} ($M=15.92, SD=3.80$) led to greater mental workload than \textit{Symbol} ($M=7.25, SD=5.36, p = .002$), \textit{Abstract Light} ($M=7.00, SD=4.00, p<.001$), and \textit{Text} ($M=9.17, SD=4.93, p<.001$). No other significant results were found.

\textbf{Necessity, Reasonability, and Preference Ranking}. Participants stated eHMI designs are necessary ($M=5.58, SD=1.24$) and reasonable ($M=5.50, SD=0.90$). For ranking data, \textit{Abstract Light} ($M=2.25, SD=1.22$) was rated the best, followed by \textit{Text} ($M=2.50, SD=1.00$) and \textit{Symbol} ($M=2.58, SD=1.51$). \textit{Situational Awareness} is the least preferred ($M=5.50, SD=0.52$), with \textit{Anthropomorphic} rated the second worst ($M=4.75, SD=1.29$). \textit{Road-based Projection} ($M=3.42, SD=1.62$) had the third-worst rating. Friedman's Test showed that there was a significant difference in the ranking data among the eHMI concepts ($\chi^2(5)=30.762, p <.001$). Post hoc analysis with Bonferroni correction showed that \textit{Situational Awareness} was rated significantly worse than \textit{Symbol} ($p=.045$), \textit{Abstract Light} ($p=.03$), and \textit{Text} ($p=.03$). In addition, \textit{Text} is rated significantly better than \textit{Anthropomorphic} ($p=.03$).

In summary, our quantitative data (questionnaire ratings and ranking) suggested \textit{Abstract Light} was the best perceived visual design, followed by \textit{Text} and \textit{Symbol}. Overall, the results appeared to align with results reported in prior works by~\citep{scalability_Mark, Dey_scalable}, except for the \textit{Road-Projection}.

\subsubsection{High-Level eHMI Design Requirements}\label{designRequirement}
For the qualitative data analysis, we conducted a thematic analysis \cite{Braun01012006} using an inductive coding approach. The interview transcripts (auto-transcribed by Teams and corrected by an author) were independently read by two coders, who familiarised themselves with the data and generated initial codes. The coders then met to organise and consolidate these codes into preliminary themes and sub-themes. Through a recursive review process, the coders refined and evaluated the emerging themes against both the coded excerpts and the full dataset. Once the themes were clearly defined and sufficiently concise, we produced the final report--the four high-level eHMI design requirements derived from this analysis, which was also used to inform the design of the VR user study:

\begin{enumerate}
  \item \textbf{Key Vehicle States and State-Transition Must be Clearly Presented}. Both groups mentioned that there is a need to display key vehicle states and a clear state-transition design from slowing down to fully stopped, so that pedestrians can see the intention and state of the AV. For instance, H10 said: "\textit{Let people see that there is a difference between the two states}." 

  \item \textbf{Avoid Crossing Advice, Let Pedestrians Decide}. Our participants raised concerns that relying on cross advice from AVs could create liability and legibility issues. This is inline with previous works by \citet{status_zhang,FAAS2020171} and Volvo\footnote{\href{https://www.media.volvocars.com/global/en-gb/media/pressreleases/237019/volvo-360c-concept-calls-for-universal-safety-standard-for-autonomous-car-communication1\#}{Volvo: Volvo 360c Concept}}. Instead of giving crossing advice, they suggest that "\textit{The choice has to be down to the pedestrian}" [D3]. Therefore, in line with works around non-disabled people~\citep{status_zhang,FAAS2020171} and suggestion given by ISO technical report~\citep{iso23049}, eHMIs should present the AV's intention/state. 

  \item \textbf{Provide Multi-Modal eHMI}. Participants from both groups first acknowledged the effectiveness of visual eHMI designs, then explicitly mentioned that eHMI should be inclusive for them and others. They expressed the need for wider accessibility and inclusiveness, such as (1) ensuring the eHMI designs are accessible under different weather and light conditions and (2) making eHMI feasible for low vision users, colour-blind people, and people with multiple disabilities. One participant explained that "\textit{for visual impaired people, visual might not be the best solution. They would probably want an audible cue as well.}" Therefore, reflecting the recent works on multi-modal eHMIs and their benefits for being inclusive and accessible to a wider audience~\citep{multi_modal_dey,Asha_wheelchair,Mark_Vision,mark_include_impairment}, there seems to be a need to develop multi-modal eHMIs.
  
  \item \textbf{Provide Combined-Visual Communications}. Not just requesting multi-modal eHMI, participants suggested that the AV should employ multiple types of visual communications so that the pedestrians could pick up the visual eHMI that works for them (e.g., dyslexic people could refer to the \textit{Abstract Light} signal, colour blind people could rely on the \textit{Text}/\textit{Symbol}) and rely on one visual eHMI if the other were damaged. "\textit{There needs to be some kind of design that don't show colour alone. You'd need some kind of other visual together}." [H11].
\end{enumerate}

In conclusion, our focus group findings suggested that \textit{Abstract Light}, \textit{Text}, and \textit{Symbol} were the top three visual eHMI candidates and should be further evaluated in our VR user study. As multi-modal eHMI was highly recommended by our focus group participants, we briefly explored the use of auditory eHMI in our VR user study. As this formative study focuses on visual eHMI designs, we conclude the choice of the auditory eHMI to be \textit{Speech} as its usefulness for other disabled people~\citep{Mark_Vision}, which met the requirement of our participants (i.e., eHMI should be inclusive for DHH people and others). 

We adjusted our designs: (1) display a clear state-transition of eHMI when the vehicle is slowing down and a clear state when the vehicle is fully stopped, and (2) avoid using instructive words or symbols. We proposed eHMI candidates that are (1) multi-modal eHMIs via audio-visual eHMI (Visual: \textit{Abstract Light}, \textit{Text}, and \textit{Symbol}; Audio: \textit{Speech}), and (2) single visual eHMIs (\textit{Abstract Light} alone) or combined-visual eHMIs (\textit{Abstract Light + Text} and \textit{Abstract Light + Symbol}) to understand how DHH people perceive them. 

\section{VR User Study}
This study was guided by three research questions (RQ):

RQ1: How do the ratings for experience (i.e., trust, acceptance, perceived safety, mental load) and behaviour (i.e., gaze behaviour, step-in road time, early step into the road count) differ between participants in the Hearing group and in the DHH group?

RQ2: What impact do the visual eHMIs have on pedestrians regarding experience and behaviour? 

RQ3: What impact does providing auditory speech-based eHMIs have on pedestrians regarding experience and behaviour? 

\begin{figure*}
 \centering
 \includegraphics[width=1\textwidth]{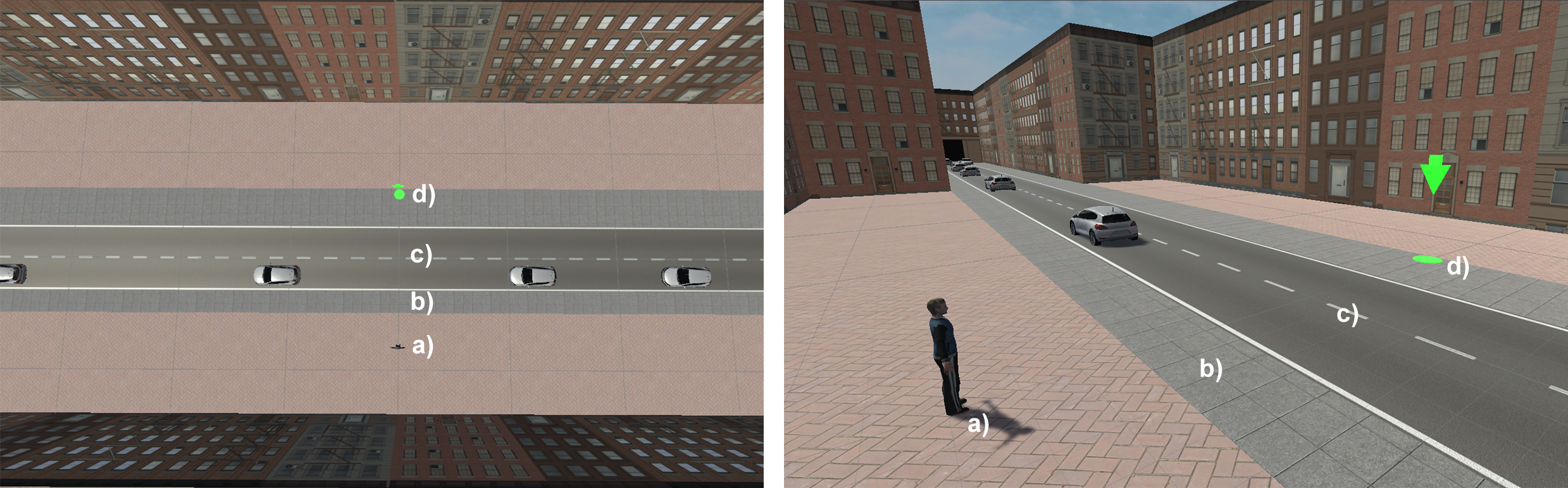}
 \caption{Setup of the virtual environment: The left shows a top-down view, while the right shows a leftward view. a) The starting position of the participant, which is 4.7 m from the road. b) The 2 m wide grey pavement area, which is used to inform the vehicle of the intention of crossing. c) The 6.8 m wide two-lane road that the participant needed to cross. d) The waypoint (green surface with downward arrow indicated), which is 2.5 m away from the road, is used to proceed with the partition to the next repetition. The human model was disabled in the experiment, placed here as a visual illusion.}
 \label{fig:environment}
 \Description{A figure shows the set up of the virtual environment with key measurements of each part of the environmental: a) The starting position of the participant, which is 4.7 m from the road. b) The 2 m wide grey pavement area, which is used to inform the vehicle of the intention of crossing. c) The 6.8 m wide two-lane road that the participant needed to cross. d) The waypoint (green surface with downward arrow indicated), which is 2.5 m away from the road, is used to proceed with the partition to the next repetition.}
\end{figure*}

\subsection{Virtual Environment}
We used Unity V2022.3.44 to develop the virtual environment. \autoref{fig:environment} illustrates a western-style city environment set up with a straight 2-way 1-lane road used in our study. Vehicles enter the area through a tunnel on one side and exit through another on the other side. Participants needed to start the trial about 4.7 m away from the first lane (see \autoref{fig:environment}a), arrive and wait at the grey pavement (see \autoref{fig:environment}b), and when there was enough gap or vehicles started to stop or stopped for the participant, they would need to cross the road (see \autoref{fig:environment}c) to the other end, until they reach the green waypoint (see \autoref{fig:environment}d), which would proceed them to the subsequent trial or finished the condition if it was the 2nd trial. 

For the first lane, the one closer to the participant, there were constantly 10 vehicles, consisting of 1 AV and 9 manual vehicles. For the second lane, we set 2 manual vehicles to drive towards the right-hand side tunnel when launching the environment; these 2 vehicles would disappear and not reappear after entering the tunnel. This setting for the second lane was to raise awareness among the participants of vehicles potentially approaching from the left-hand side. At the same time, we do not want the vehicles in the second lane to affect the crossing decision and behaviour of the participants. The maximum speed of the city environment was set at 50 km/h, while vehicles drive from 40 km/h to 50 km/h. The participant will cross the road from a) to d) twice. Only the AV would stop for the participants; however, we informed the participants that both manual vehicles and AVs may stop for them. The simulation also provided a light ambient background noise. 

\begin{figure}[t]
 \centering
 \includegraphics{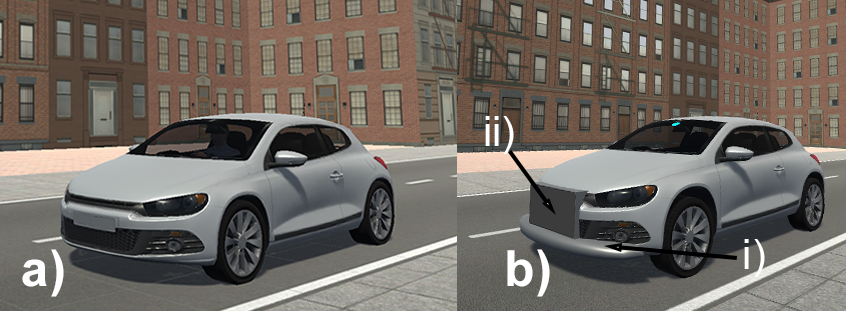}
 \caption{a) The Manual vehicle and b) the AV used in the study. i) The Light Strip, ii) The Display.}
 \label{fig:manual_AV}
 \Description{The visual appearance of the manual vehicle on the left and the visual appearance of the AV on the right.}
\end{figure}

~\autoref{fig:manual_AV} shows the appearance of the manual and AVs used in the user study. The AV used the same vehicle model as the manual vehicle but with three modifications in line with the prior work~\cite{FAAS2020171,Mahadevan_audio,scalability_Mark}: (1) a round cyan light positioned at the top centre of the vehicle’s windshield to indicate that the vehicle is driving autonomously, (2) a light strip located at the bottom of the bumper region to display the light design, and (3) a display located at the grill region of the bumper to display the text or symbol. To further differentiate the AV and the manual vehicle, the manual vehicle features a human driver character inside the vehicle, while the AV does not involve any human character inside.

\subsection{User Study Design}
\begin{figure*}[t]
 \centering
 \includegraphics[width=1\textwidth]{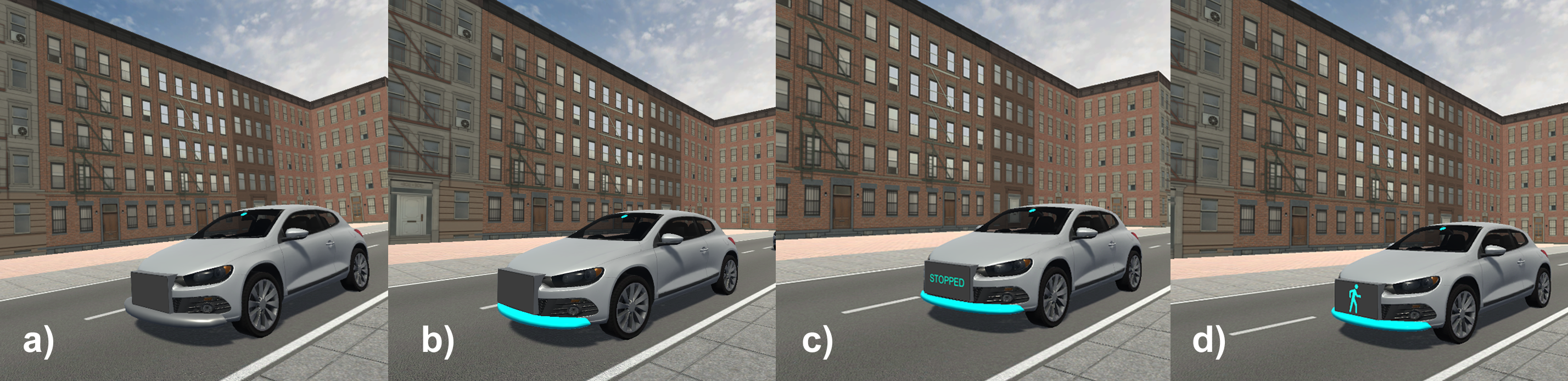}
 \caption{The visual eHMI conditions used in the user study, showing when the vehicle is fully stopped: a) \textit{No Visual}: No added visual effects; b) \textit{Abstract Light}: the light strip stayed on; c) \textit{Text}: the light strip stayed on while the display shows "STOPPED"; d) \textit{Symbol}: the light strip stayed on while the display shows a symbol of "Walking Man".}
 \label{fig:visual_ehmi}
 \Description{A figure shows 4 visual eHMI design used in the study when the autonomous vehicle is fully stopped: a) No Visual: No added visual effects; b) Abstract Light: the light strip stayed on; c) Text: the light strip stayed on while the display shows "STOPPED"; d) Symbol: the light strip stayed on while the display shows a symbol of "Walking Man".}
\end{figure*}

We employed a 4 $\times$ 2 within-subjects design with Visual eHMI and Auditory eHMI as the independent variables. The 4 levels involved in the Visual eHMI (see Fig. \ref{fig:visual_ehmi}) are: 

\begin{enumerate}
\item \textit{No Visual}. This condition serves as the baseline condition; both the light strip and the display would be presented but not activated; we want to ensure that differences measured between the eHMI concepts solely rely on the eHMI concept itself. 

\item \textit{Abstract Light}. While the AV slows down the process, the light strip pulses in Cyan and the frequency goes from fast (pulsates between on and off at a rate of 1 Hz) to slow (pulsates between on and off at a rate of 0.5 Hz) to indicate a speed change. When fully stopped, the light strip would stop pulsing but remain static. Unless the vehicle started to yield for participants, neither the light strip nor the display would show any additional information. This would save energy consumption and make the vehicle more environmentally friendly. 

\item \textit{Abstract Light + Text}. The implementation of this condition is analogous to \textit{Abstract Light}. The main difference is that the display would show "SLOWING" during the slowdown process and show "STOPPED" when the AV fully stopped. Both words were presented in bold cyan letters.

\item \textit{Abstract Light + Symbol}. The implementation of this condition is analogous to \textit{Abstract Light}. The main difference is that the display would show the "Standing Man" symbol during the slowdown process and the "Walking Man" symbol when the AV fully stopped, both symbols presented in colour cyan.
\end{enumerate}

The 2 Auditory levels are: 
\begin{enumerate}
\item \textit{Without Speech}. This condition serves as the baseline condition with no added speech eHMI. We assumed all vehicles would be electric vehicles; therefore, we embedded the Acoustic Vehicle Alerting System (AVAS) sound from BMW into all vehicles. The AVAS had a volume of 65 dB. 

\item \textit{With Speech}. On top of the above condition, the AV would play a verbal message "I'm deaccelerating" after 1.5 s when the vehicle started to slow down. The length of the message is 1.3 s. When the vehicle fully stopped, it would immediately play a verbal message "I'm stopped", the length of the verbal message is 0.8 s. Both messages only played once. We used a free online Text-to-Voice service\footnote{\href{https://ttsmp3.com/}{TTSMP3 Text-to-Speech services; setting: US English / Kendra}} for generating the message.
\end{enumerate}

The order of Visual eHMI $\times$ Auditory conditions was counterbalanced in this study.

\subsection{Apparatus and Setup}
We used a Varjo XR-4 focal edition device as our VR headset, which offers a 90 Hz refresh rate with 3840 $\times$ 3744 resolution and a 120° $\times$ 105° field of view. We also used the Varjo XR-4 built-in eye tracker, which offers 200 Hz eye tracking and provides gaze visualisation, video recording, and eye measurements such as pupil iris diameter, openness, and interpupillary distance~\footnote{\href{https://developer.varjo.com/docs/unity-xr-sdk/eye-tracking-with-varjo-xr-plugin}{Varjo Developer Eye Tracking}}. To provide an immersive experience, we used Cyberith Virtualizer Elite 2 as our walking simulation for locomotion. Both devices were connected to a PC with an i9 CPU, 64 GB RAM, and a GeForce RTX 4090 Graphics card. Sound was provided by the VR device directly and the sound loudness level on the PC setting for Varjo headset was set to the same level across all participants; we did not employ any additional headphones or earphones, as they would clash with hearing aids or implants use. The study was conducted in a well-illuminated indoor laboratory.

\subsection{Outcome Measurement}
We used both subjective and objective measurements: 

\subsubsection{Crossing Experience}
We used the same set of questionnaires (Trust in Automation~\cite{trust_korber}, van der Laan acceptance scale~\cite{VANDERLAAN19971}, Perceived Safety~\cite{safety_Faas}) as we used in the Formative Study, expect the NASA-TLX question for Mental Demand. Instead, we employed the Low/High Index of Pupillary Activity ~(LHIPA)~\cite{LHIPA} for measuring the workload calculated based on the pupil diameter data provided by the Varjo-XR 4 device. 

\subsubsection{Crossing Behaviour}
This was evaluated by Eye Behaviour data and Movement data. For Eye Gaze Behaviour, we recorded the number of fixations among three areas of interest (AOI) on the AV--(1) the Light Strip, (2) the display, (3) the other parts of the AV. We reported the following 5 data metrics: (1) \textbf{Light Strip Duration}: Fixation on the Light Strip measured in second, (2) \textbf{Display Duration}: Fixation on the Display measured in second, (3) \textbf{Whole Vehicle Duration}: Fixation on the combination of Light Strip, Display, and other parts of Vehicle, measured in second (4) \textbf{Active Visual eHMI Duration}: duration on active visual eHMI for conditions that employ Visual eHMI, i.e., the Light Strip for \textit{Abstract Light}, combination of the Light Strip and the Display for both \textit{Abstract Light + Text} and \textit{Abstract Light + Symbol} conditions, and (5) \textbf{Active Visual eHMI Percentage}: (Active Visual eHMI Duration $/$ the Whole Vehicle Duration) $\times$ 100\%.

We counted the time when the AV was 20 m away from the participant's crossing point till the participant stepped into the road. We choose 20 m as the starting point for us to calculate the results, because it provides higher accuracy, and early research suggests that pedestrians started to look more at the vehicle than the road ahead of the vehicle from this range~\citep{10.1145/3342197.3344523}.

\textit{Step Into the Road Time} and \textit{Early Step Into the Road Count} were reported as our Movement data. \textit{Step Into the Road Time}: The time taken by the participant to start crossing the road from the moment an AV starts to slow down on the nearest lane. \textit{Early Step Into the Road Count}: The number of times the participant stepped onto the road before the vehicle fully stopped.

\subsubsection{Post-Study Questionnaire}
\textit{Necessity and Reasonability}: Following~\cite{scalability_Mark}, we employed a 7-point (1=Totally Disagree to 7=Totally Agree) Likert scale to measure participants' perception of the Necessity and Reasonability of (1) visual and (2) auditory eHMI concepts. \textit{Rank}: We also collected the ranking of the visual eHMI concepts and audio conditions; the lower the number, the better. 

\subsubsection{Semi-structured Interview}
We first asked the participants to reflect on their design-making strategy:"Can you reflect on how you decided to cross in front of the approaching AV?" Then the interview moved to each visual eHMI design used in the studies (i.e., \textit{Abstract Light}, \textit{Abstract Light + Text}, and \textit{Abstract Light + Symbol}), we asked three questions "Overall, what do you think about the [condition name]", "What did you like about this condition?", "What did you not like about this condition?". For auditory, we asked "Overall, what do you think about the verbal feedback?", "What did you like about it?", "What did you not like about it?", "Have the hearing levels impacted your communication with the autonomous vehicle? Were you able to hear it?". In the end, we offer an open question to ask if participants have anything to add.

\subsection{Participants}
Participants were recruited via physical posters, social media platforms, charities, and informal referrals through word-of-mouth. Thirty-two participants took part in the study. 16 (\textit{Mean} $age = 25.63, SD=5.82$; 10 women; 6 men) participants self-identified as having average hearing for both ears. The other 16 (\textit{Mean} $age = 45.56, SD=19.49$; 10 women; 6 men) participants self-identified as DHH people; full details can be found in \autoref{table:VR_DHH}. A BSL interpreter was presented in-person to assist with the study for BSL users. 

All participants were instructed to cross as if they were in real life. For DHH participants, this meant they had the option to wear or not wear their hearing aids or implants, depending on their daily experience. Thirteen DHH participants who used hearing aids and implant daily decided to use it in this study as this is their everyday experience, 1 Deaf participant (i.e., P26) who used implant but only for indoor environment decided not to use it in this study as this is not usual for him, and 2 HoH participants were not using hearing aids at all and did not have one. DHH participants who used hearing aids or implants in the study were instructed to put it to their usual mode when they walk on the street.

On a 5-point Likert scale (1 = Not at all, 5 = Definitely), Hearing participants showed slight interest in AVs ($M=2.69, SD=1.20$) and had some knowledge about AV ($M=4.13, SD=1.02$). Similar results were found among DHH participants, while ratings for the interests in AVs had an average score of 2.63 ($SD=1.45$), and the ratings for their knowledge in AV were 3.88 ($SD=0.96$).

\begin{table}
\caption{Demographic information of DHH participants in the VR user study}
\begin{tabular}{p{0.03\linewidth} p{0.04\linewidth} p{0.08\linewidth} p{0.14\linewidth} p{0.14\linewidth} p{0.13\linewidth} p{0.15\linewidth}}
\hline
    ID & Age & Gender & Left Ear HL Level & Right Ear HL Level& Preferred Communication & Use Hearing Technology?\\
    \hline
    P17 & 77 & Male & Severe & Severe & English &Yes  \\
    P18 & 70 & Male & Moderate & Moderate & English  & Yes  \\
    P19 & 23 & Female & Mild & Normal & English & No  \\
    P20 & 66 & Female & Mild & Mild & English & Yes  \\
    P21 & 59 & Female & Profound & Severe & English & Yes  \\
    P22 & 69 & Female & Moderate & Severe & English & Yes  \\
    P23 & 28 & Female & Moderate & Normal & English & Yes  \\
    P24 & 28 & Female & Mild & Normal & English & No  \\
    P25 & 27 & Male	& Profound & Severe & BSL & Yes\\
    P26 & 25 & Female & Profound & Profound & BSL & No\\
    P27 & 21 & Male & Profound & Profound & English & Yes  \\
    P28 & 49 & Female & Moderate & Mild & English & Yes  \\
    P29 & 33 & Female & Profound & Profound & BSL & Yes  \\
    P30 & 41 & Male	& Profound& Profound& BSL	& Yes\\
    P31 & 61 & Male & Moderate & Severe & English & Yes \\
    P32 & 52 & Female & Moderate & Profound & English & Yes\\
    \hline
    \end{tabular}
    \footnotesize 
    Hearing Loss (HL)
  \label{table:VR_DHH}
  \Description{Demographic information of DHH participants in our VR user study, listing with their ID, Age, Gender, left and right ear hearing loss level, preferred mode of communication, and whether or not they use hearing technology.}
\end{table}

\subsection{Procedure}
Each study started with a brief introduction, the consent form to review and sign, and a demographic questionnaire. Before the formal study, participants would wear the Varjo XR-4 and undergo two trials of road crossing in the same environment but without any vehicles present to familiarise themselves with the Virtualizer Elite 2. Once they were ready, they would proceed to the formal experiment conditions. In each condition, participants had to cross the street twice. Eye calibration was checked through Varjo Base software at the beginning of each condition. After each condition, participants answered the required questionnaires. At the end of the study, they completed a post-study questionnaire and then participated in a semi-structured interview. 

\section{Results}
Shapiro-Wilk suggested that all data were not normally distributed, so we applied the Aligned Rank Transform (ART) to the data~\cite{ART_ANOVA}. For questionnaire data, we employed a three-way mixed ANOVA with Visual (\textit{No Visual}, \textit{Abstract Light}, \textit{Abstract Light + Text}, and \textit{Abstract Light + Symbol}) and Audio (\textit{With Speech} and \textit{Without Speech}) as within-subjects variable and Group (1. Hearing and 2. DHH) as the between-subjects variable. For necessity and reasonability of visual and auditory eHMIs, we employed a two-way mixed ANOVA after transforming the data with the ART method; the within-subject factor was Modality (Visual and Auditory), while the between-subject factor was the Group (Hearing and DHH). Post-hoc tests were conducted using the ART-C method~\cite{ART_posthoc} with Bonferroni corrections.

\subsection{Crossing Experience}

\subsubsection{Trust in Automation} 
\begin{figure*}[t]
    \centering
    \begin{minipage}{0.49\textwidth}
        \centering
        \includegraphics[width=\linewidth]{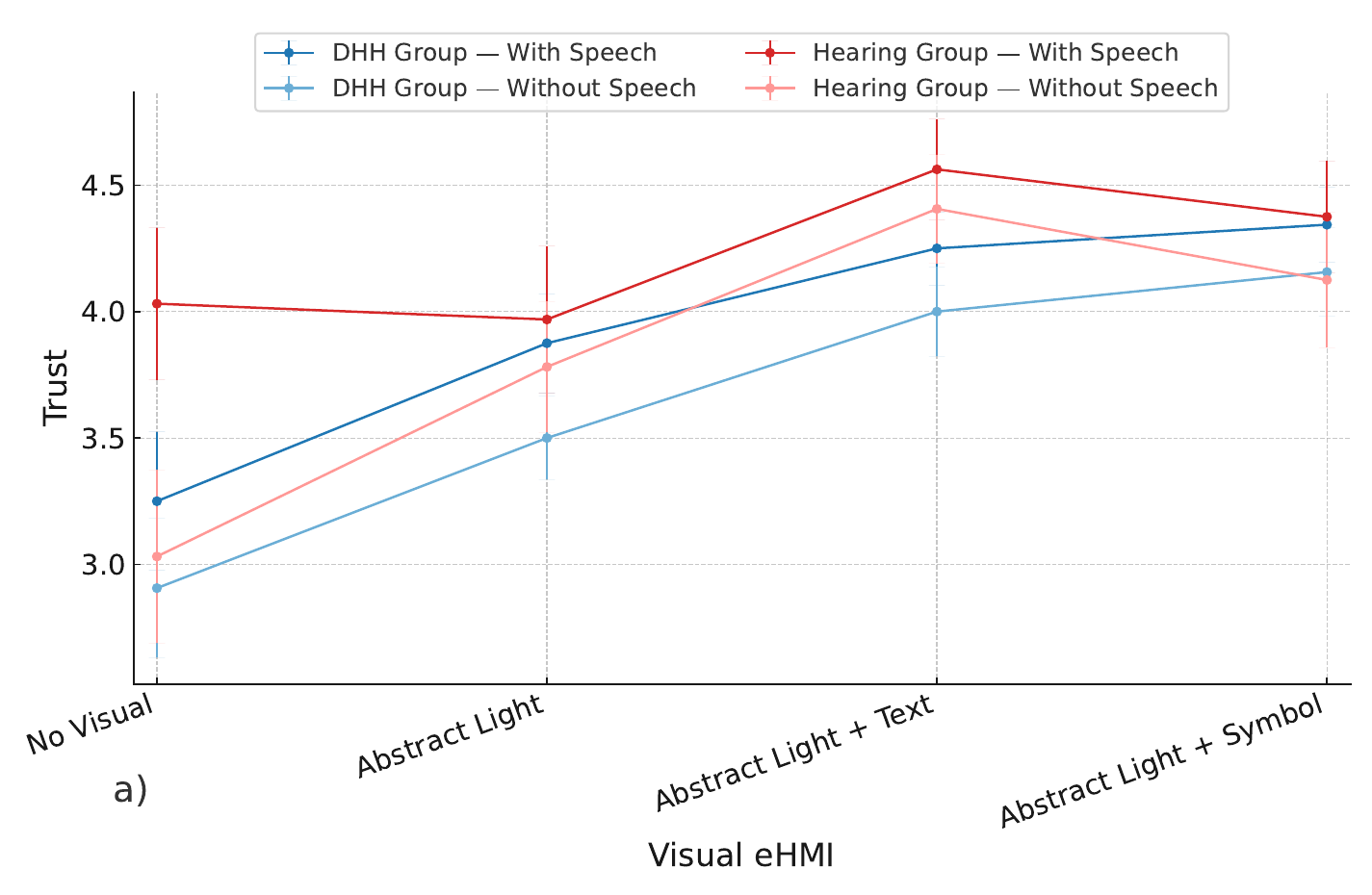}
    \end{minipage}
    \hfill
    \begin{minipage}{0.49\textwidth}
        \centering
        \includegraphics[width=\linewidth]{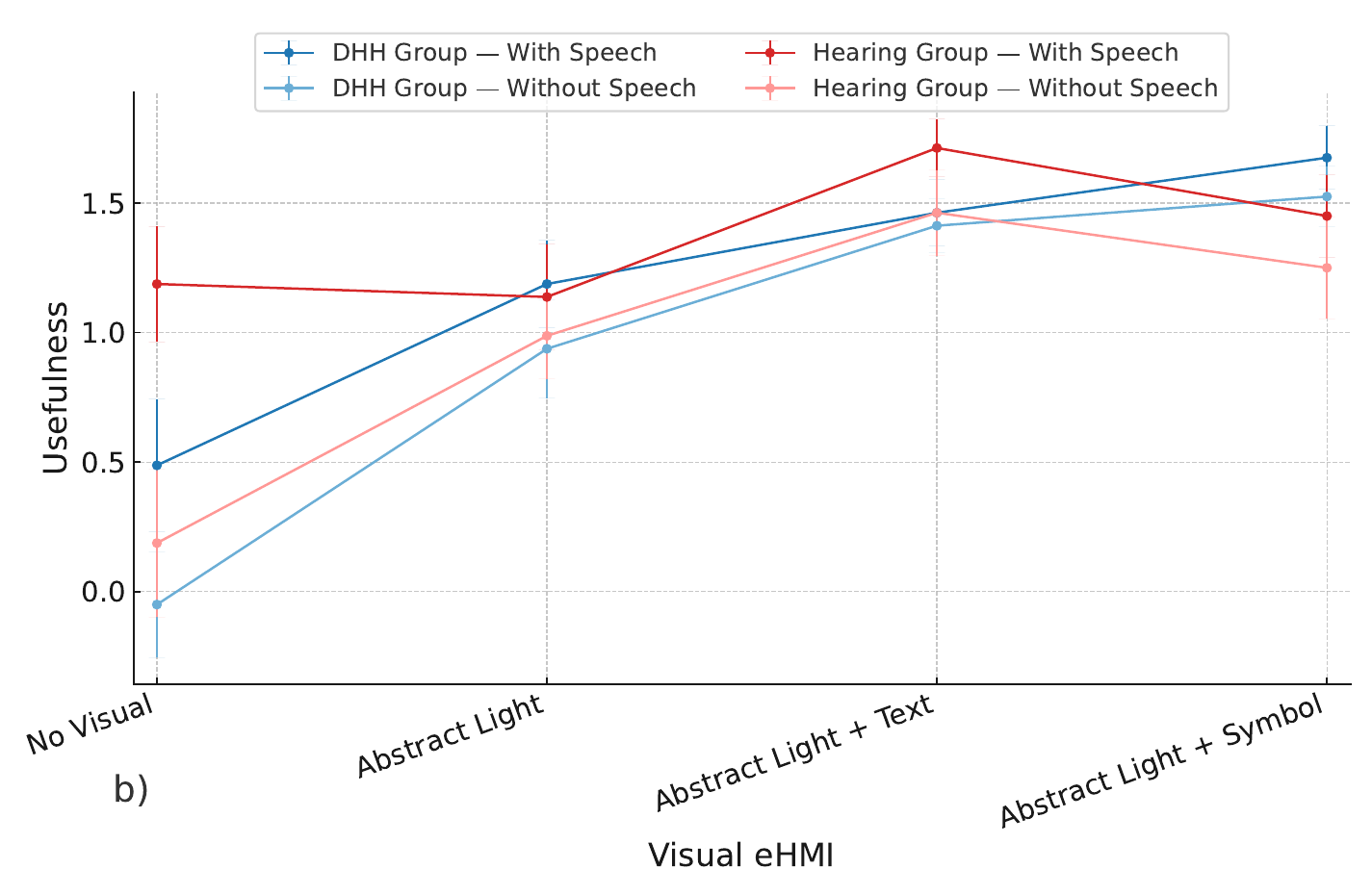}
    \end{minipage}
    \caption{a) Trust and b) Usefulness ratings by Visual × Audio eHMI conditions, separated for DHH and Hearing groups, with and without speech. Error bars show $\pm$SE.}
    \Description{Data figure showing a) trust and b) usefulness ratings by Visual and Audio eHMI conditions for the DHH group.}
    \label{fig:trust_usefulness}
\end{figure*}

Regarding Understandability, we could not observe any significant effect. As for Trust subscale, ANOVA results showed a significant main effect of Visual ($F(3,210) = 26.403, p < .001, \eta_{p}^{2}=.274$) and Audio ($F(1,210) = 20.253, p < .001, \eta_{p}^{2}=.088$) on Trust. Post-hoc pairwise comparisons revealed that the \textit{Abstract Light + Symbol} ($M=4.25, SD=0.82$) and the \textit{Abstract Light + Text} ($M=4.30, SD=0.75$) had significantly higher Trust ratings than the \textit{Abstract Light} ($M=3.78, SD=0.93, p<.001$) and the \textit{No Visual} ($M=3.30, SD=1.25, p<.001$). In addition, the \textit{Abstract Light} ($M=3.78, SD=0.93$) had a better Trust rating than the \textit{No Visual} ($M=3.30, SD=1.25, p=.025$). Post-hoc pairwise comparisons on the main effect Audio showed that \textit{With Speech} condition ($M=4.08, SD=0.97$) led to significantly better Trust ratings than \textit{Without Speech} condition ($M=3.74, SD=1.07, p<.001$). We did not find any interaction effect. Mean Trust ratings among each condition can be found in \autoref{fig:trust_usefulness}a.

\subsubsection{Acceptance} Usefulness. Mean Usefulness ratings among each condition across participants in Hearing and DHH groups can be found in \autoref{fig:trust_usefulness}b. ANOVA tests yielded significant main effects of Visual ($F(3,210) = 42.088, p < .001, \eta_{p}^{2}=.375$) and Audio ($F(1,210) = 27.10108, p < .001, \eta_{p}^{2}=.114$). Post-hoc pairwise comparisons for the main effect Visual showed that both the \textit{Abstract Light + Symbol} ($M=1.48, SD=0.61$) and the \textit{Abstract Light + Text} ($M=1.51, SD=0.52$) had significantly higher Usefulness ratings than the \textit{Abstract Light} ($M=1.06, SD=0.72, p<.001$) and the \textit{No Visual} ($M=0.45, SD=1.06, p<.001$). In addition, the \textit{Abstract Light} ($M=1.06, SD=0.72$) had a better Usefulness rating than the \textit{No Visual} ($M=0.45, SD=1.06, p<.001$). Post-hoc pairwise comparisons for the main effect Audio showed that \textit{With Speech} condition ($M=1.29, SD=0.78$) resulted in a significantly higher Usefulness rating than \textit{Without Speech} condition ($M=0.96, SD=0.92, p<.001$). 

We observed an interaction effect of Visual $\times$ Group ($F(3,210) = 4.617, p = .004, \eta_{p}^{2}=.062$) on Usefulness. Post-hoc pairwise comparisons results among the participants in the Hearing group showed that the \textit{Abstract Light + Symbol} ($M=1.35, SD=0.71$) was significantly better than the \textit{No Visual} ($M=0.69, SD=1.13, p<.001$), the \textit{Abstract Light + Text} ($M=1.59, SD=0.57$) was significantly better than the \textit{Abstract Light} ($M=1.06, SD=0.74, p=.001$) and the \textit{No Visual} ($M=0.69, SD=1.13, p<.001$). Post-hoc pairwise comparisons results among the DHH participants showed that the \textit{Abstract Light + Symbol} ($M=1.60, SD=0.48$) had a significantly higher Usefulness rating than the \textit{Abstract Light} ($M=1.06, SD=0.72, p<.001$) and the \textit{No Visual} ($M=0.22, SD=0.95, p<.001$). The \textit{Abstract Light + Text} ($M=1.44, SD=0.46$) and the \textit{Abstract Light} ($M=1.06, SD=0.72$) were both significantly rated higher than the \textit{No Visual} ($M=0.22, SD=0.95$, both $p<.001$).

We also observed an interaction effect of Visual $\times$ Audio ($F(3,210) = 5.157, p = .002, \eta_{p}^{2}=.069$) on Usefulness. Post-hoc pairwise comparisons results among the \textit{No Visual} condition showed that \textit{With Speech} condition ($M=0.84, SD=1.01$) resulted in a higher Usefulness score than \textit{Without Speech} condition ($M=0.07, SD=0.98, p=.003$). Post-hoc pairwise comparisons results also showed that when Speech was provided (1) the \textit{Abstract Light + Symbol} ($M=1.56, SD=0.57$) was significantly better than the \textit{Abstract Light} ($M=1.16, SD=0.74, p=.042$) and the \textit{No Visual} ($M=0.84, SD=1.01, p<.001$), (2) the \textit{Abstract Light + Text} ($M=1.59, SD=0.49$) was significantly better than the \textit{No Visual} ($M=0.84, SD=1.01, p<.001$). Post-hoc pairwise comparisons results showed that when Speech was not provided, (1) the \textit{Abstract Light + Symbol} ($M=1.39, SD=0.65$) had a significantly higher score than the \textit{Abstract Light} ($M=0.96, SD=0.70, p=.035$) and the \textit{No Visual} ($M=0.07, SD=0.98, p<.001$), (2) the \textit{Abstract Light + Text} ($M=1.44, SD=0.55$) had a significantly higher score than the \textit{Abstract Light} ($M=0.96, SD=0.70, p=.007$) and the \textit{No Visual} ($M=0.07, SD=0.98, p<.001$), and (3) the \textit{Abstract Light} ($M=0.96, SD=0.70$) had a significantly higher score than the \textit{No Visual} ($M=0.07, SD=0.98, p=.003$).

\begin{figure*}[t]
    \centering
    \begin{minipage}{0.49\textwidth}
        \centering
        \includegraphics[width=\linewidth]{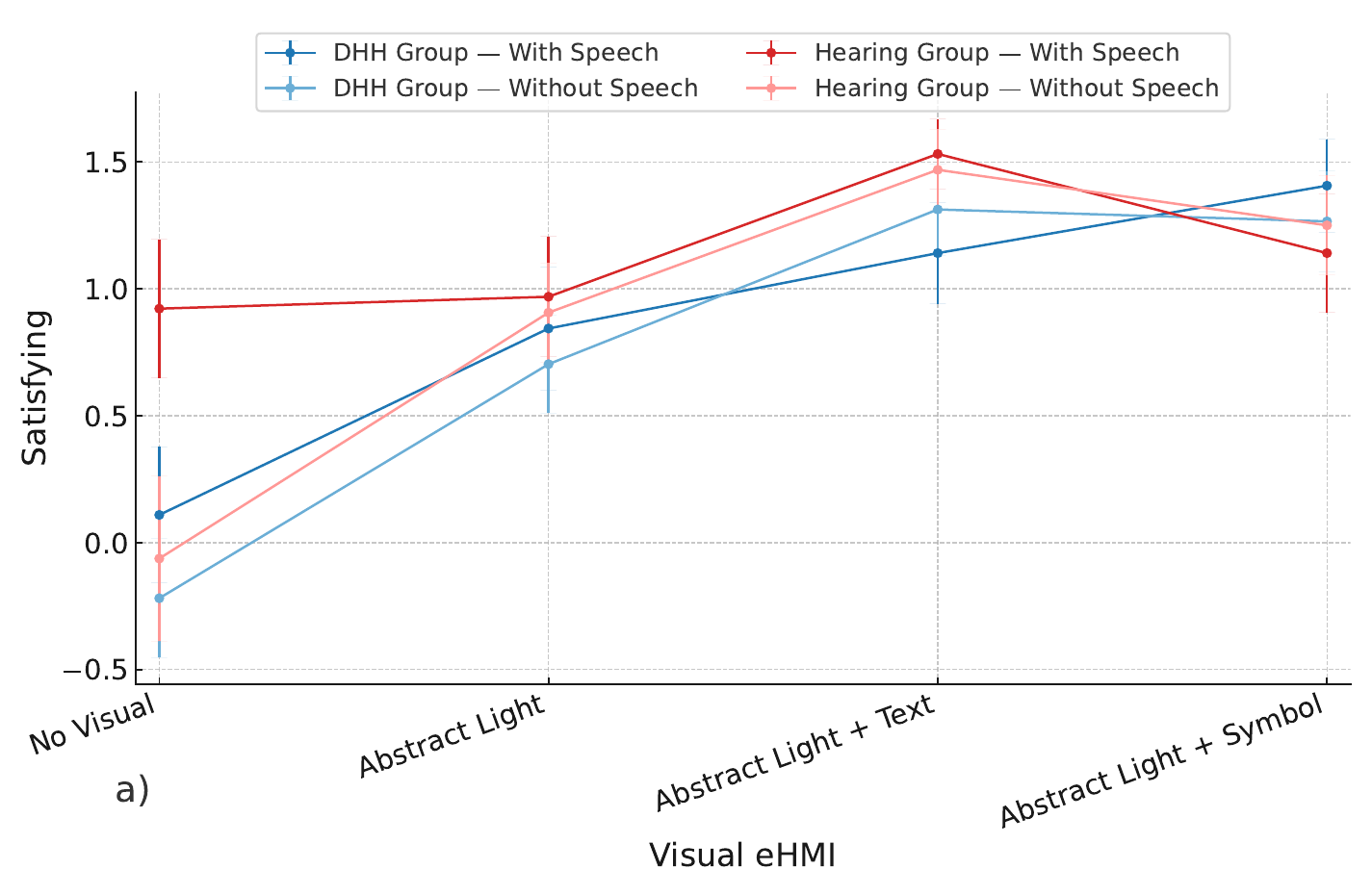}
    \end{minipage}
    \hfill
    \begin{minipage}{0.49\textwidth}
        \centering
        \includegraphics[width=\linewidth]{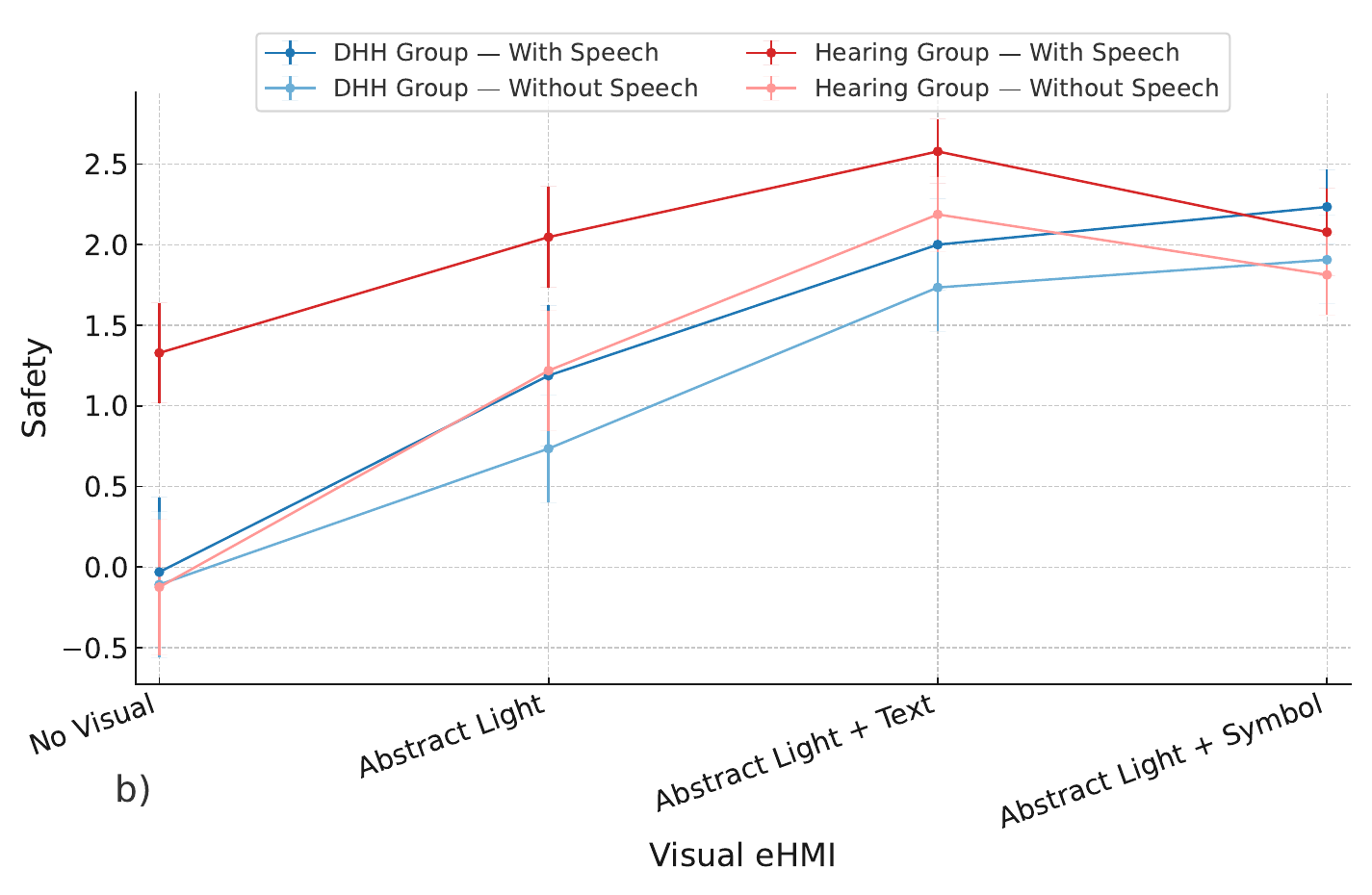}
    \end{minipage}
    \caption{a) Satisfying and b) Perceived Safety ratings by Visual × Audio eHMI conditions, separated for DHH and Hearing groups, with and without speech. Error bars show $\pm$SE.}
    \label{fig:satisfying_safety}
    \Description{Data figure showing a) satisfying and b) safety ratings by Visual and Audio eHMI conditions for the DHH group.}
\end{figure*}

Satisfying. \autoref{fig:satisfying_safety}a illustrates mean Satisfying ratings among each condition across participants in the Hearing and DHH groups. ANOVA tests yielded significant main effect of Visual ($F(3,210) = 33.807, p < .001, \eta_{p}^{2}=.326$) and Audio ($F(1,210) = 7.335, p = .007, \eta_{p}^{2}=.034$). Post-hoc pairwise comparisons for the main effect Visual showed that both the \textit{Abstract Light + Symbol} ($M=1.27, SD=0.80$) and the \textit{Abstract Light + Text} ($M=1.36, SD=0.68$) had significantly higher Satisfying ratings than the \textit{Abstract Light} ($M=0.86, SD=0.86, p<.001$) and the \textit{No Visual} ($M=0.19, SD=1.17, p<.001$), in addition, the \textit{Abstract Light} ($M=0.86, SD=0.86$) had a significantly higher Satisfying rating than the \textit{No Visual} ($M=0.19, SD=1.17, p<.001$). Post-hoc pairwise comparisons for the main effect Audio showed that \textit{With Speech} condition ($M=1.01, SD=0.97$) resulted in a significantly higher Satisfying rating than \textit{Without Speech} condition ($M=0.83, SD=1.03, p=.007$). 

We also observed an interaction effect of Visual $\times$ Audio $F(3,210) = 4.643, p = .004, \eta_{p}^{2}=.062$) on Satisfying. Post-hoc pairwise comparisons showed when Speech was provided, the \textit{Abstract Light + Symbol} ($M=1.27, SD=0.84$) and the \textit{Abstract Light + Text} ($M=1.34, SD=0.70$) were both significantly better than the \textit{No Visual} ($M=0.52, SD=1.14$, both $p<.001$). Post-hoc pairwise comparisons showed when Speech was not provided, (1) the \textit{Abstract Light + Symbol} ($M=1.26, SD=0.78$) was significantly better than the \textit{Abstract Light} ($M=0.80, SD=0.77,p=.026$) and the \textit{No Visual} ($M=-0.14, SD=1.12, p<.001$), (2) the \textit{Abstract Light + Text} ($M=1.39, SD=0.67$) was significantly better than the \textit{Abstract Light} ($M=0.80, SD=0.77,p=.002$) and the \textit{No Visual} ($M=-0.14, SD=1.12, p<.001$), and (3) the \textit{Abstract Light} ($M=0.80, SD=0.77$) was significantly better than the \textit{No Visual} ($M=-0.14, SD=1.12,p=.002$).

\subsubsection{Perceived Safety} \autoref{fig:satisfying_safety}b illustrates mean ratings among each condition across participants in the Hearing and DHH groups. ANOVA tests yielded significant main effect of Visual ($F(3,210) = 37.671, p < .001, \eta_{p}^{2}=.350$) and Audio ($F(1,210) = 21.134, p < .001, \eta_{p}^{2}=.091$) on Perceived Safety ratings. Post-hoc pairwise comparisons for the main effect Visual showed that the \textit{Abstract Light + Symbol} ($M=2.01, SD=1.02$) had a significantly higher Perceived safety rating than the \textit{Abstract Light} ($M=1.30, SD=1.51, p=.001$) and the \textit{No Visual} ($M=0.27, SD=1.74, p<.001$). Similarly, the \textit{Abstract Light + Text} ($M=2.12, SD=1.02$) had a significantly higher Perceived Safety rating than both the \textit{Abstract Light} ($M=1.30, SD=1.51, p<.001$) and the \textit{No Visual} ($M=0.27, SD=1.74, p<.001$). In addition, the \textit{Abstract Light} ($M=1.30, SD=1.51$) also had a significantly higher Perceived Safety rating than the \textit{No Visual} ($M=0.27, SD=1.74, p <.001$). Post-hoc pairwise comparisons for the main effect Audio showed that \textit{With Speech} condition ($M=1.68, SD=1.49$) resulted in a significantly higher Perceived Safety rating than \textit{Without Speech} condition ($M=1.17, SD=1.56, p<.001$). 

We also observed an interaction effect of Visual $\times$ Group ($F(3,210) = 3.162, p = .026, \eta_{p}^{2}=.043$) on Perceived Safety. Post-hoc pairwise comparisons results among the participants in the Hearing group showed that the \textit{Abstract Light + Symbol} ($M=1.95, SD=1.03$) had a significantly higher rating than the \textit{No Visual} ($M=0.60, SD=1.63, p<.001$), the \textit{Abstract Light + Text} ($M=2.38, SD=0.88$) had a significantly higher rating than the \textit{Abstract Light} ($M=1.63, SD=1.42, p=.017$) and the \textit{No Visual} ($M=0.60, SD=1.63, p<.001$). The \textit{Abstract Light} ($M=1.63, SD=1.42$) had a significantly higher rating than the \textit{No Visual} ($M=0.60, SD=1.63, p=.006$). Post-hoc pairwise comparisons results among the participants in the DHH group showed that the \textit{Abstract Light + Symbol} ($M=2.07, SD=1.01$) had a significantly higher rating than the \textit{Abstract Light} ($M=0.96, SD=1.55, p<.001$) and the \textit{No Visual} ($M=-0.07, SD=1.81, p<.001$). Similarly, the \textit{Abstract Light + Text} ($M=1.87, SD=1.10$) had a significantly higher rating than the \textit{Abstract Light} ($M=0.96, SD=1.55, p=.007$) and the \textit{No Visual} ($M=-0.07, SD=1.81, p<.001$).

\subsubsection{Workload - Ihipa}
We observed an interaction effect of Visual $\times$ Audio $F(3,210) = 2.710, p = .046, \eta_{p}^{2}=.037$. However, post-hoc pairwise comparisons showed no significant differences between conditions. No other main effects and interaction effects were found. 

\subsection{Crossing Behaviour} 

\subsubsection{Eye Behaviour} All conditions were included for the data analysis of \textbf{Light Strip Duration}, \textbf{Display Duration}, and \textbf{Whole Vehicle Duration}. For \textbf{Active Visual eHMI Duration} and \textbf{Active Visual eHMI Percentage} analysis, we discarded No Visual condition as it did not involve active eHMIs.

\begin{figure*}[t]
    \centering
    \begin{minipage}{0.49\textwidth}
        \centering
        \includegraphics[width=\linewidth]{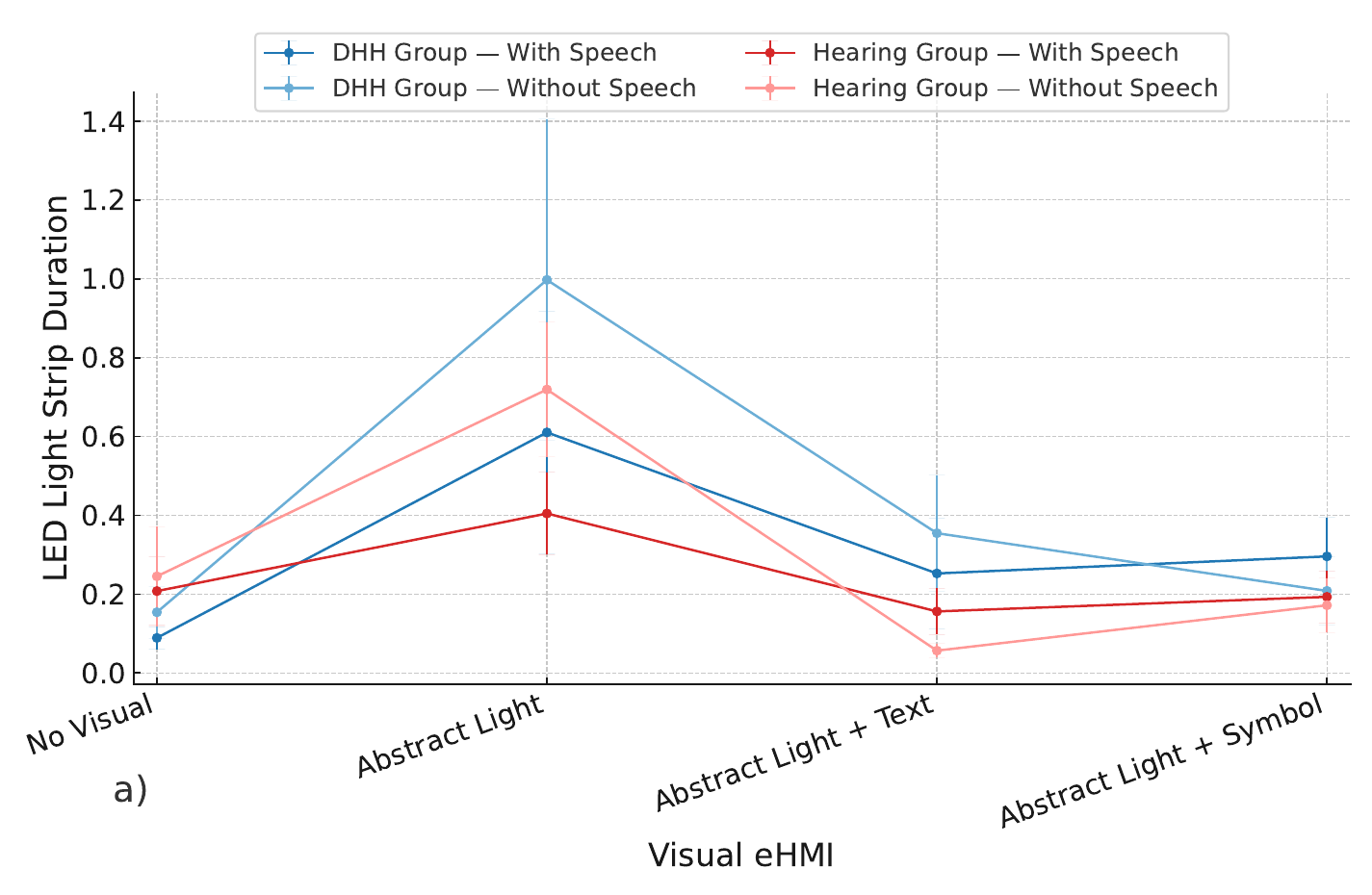}
    \end{minipage}
    \hfill
    \begin{minipage}{0.49\textwidth}
        \centering
        \includegraphics[width=\linewidth]{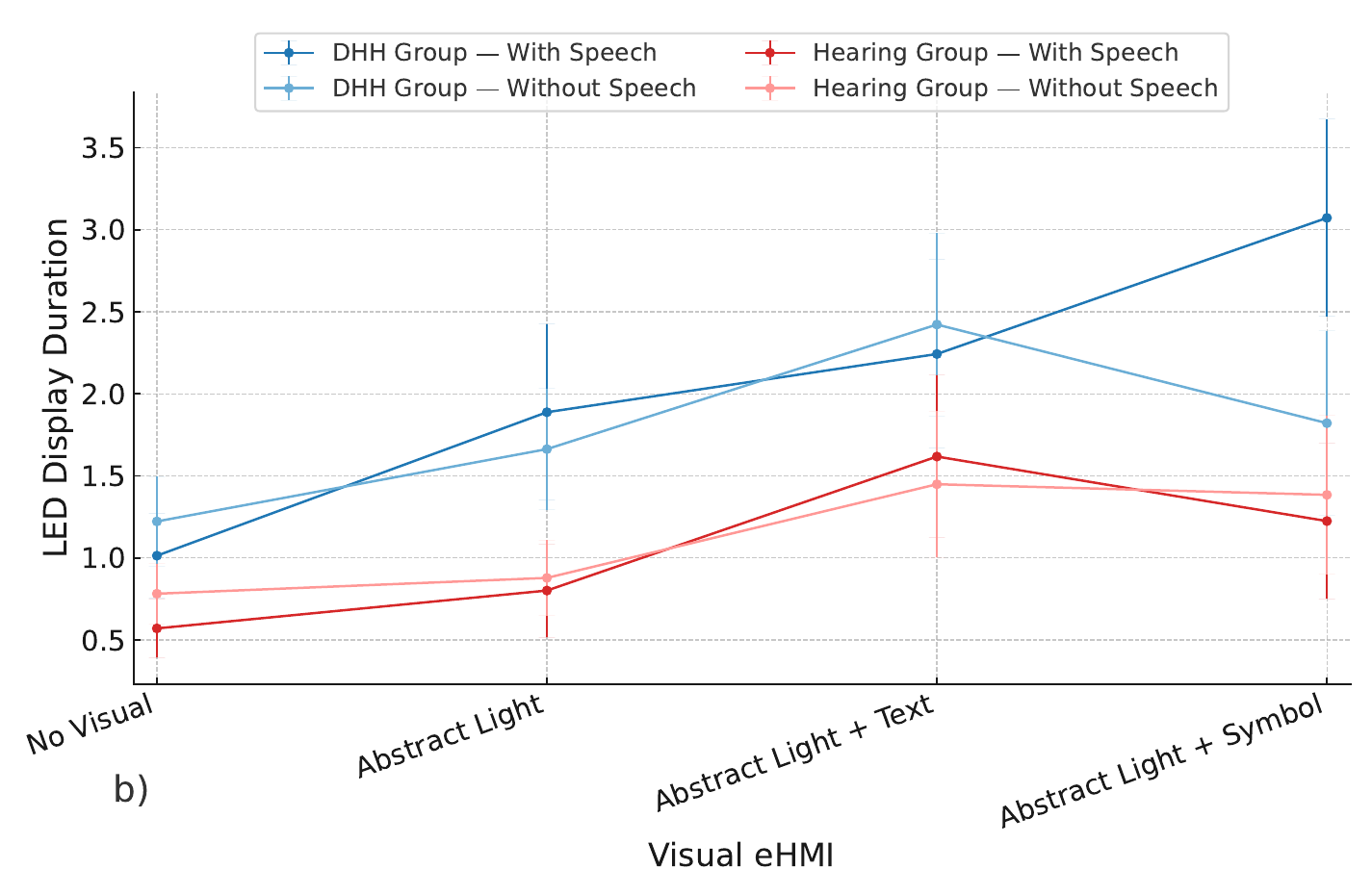}
    \end{minipage}
    \caption{a) Light Strip Duration and b) Display Duration by Visual × Audio eHMI conditions, separated for DHH and Hearing groups, with and without speech. Error bars show $\pm$SE.}
    \label{fig:BL_BD}
    \Description{Data figure showing duration spent on a) Light Strip and b) Display by Visual and Audio eHMI conditions for the DHH group.}
\end{figure*}

\textbf{Light Strip Duration}. \autoref{fig:BL_BD}a shows duration of participants in the Hearing and DHH groups spent on the Light Strip. ANOVA tests yielded significant main effect of Visual ($F(3,210) = 13.940, p < .001, \eta_{p}^{2}=.166$) and Audio ($F(1,210) = 8.341, p = .004, \eta_{p}^{2}=.038$). We also observed an interaction effect of Visual $\times$ Audio ($F(3,210) = 4.167, p = .007, \eta_{p}^{2}=.056$). We carried out post-hoc pairwise comparisons among the interaction effect, which indicated that when speech was not provided, participants looked significantly longer at the \textit{Abstract Light} ($M=0.86, SD=1.24$) than the \textit{No Visual} ($M=0.20, SD= 0.40, p<.001$), as well as the \textit{Abstract Light + Text} ($M=0.21, SD=0.44, p=.003$) and the \textit{Abstract Light + Symbol} ($M=0.19, SD=0.31, p<.001$). We did not find other significant results.

\textbf{Display Duration}. ANOVA tests yielded significant main effect of Visual ($F(3,210) = 6.163, p < .001, \eta_{p}^{2}=.081$). Post-hoc pairwise comparisons for the main effect Visual showed that participants looked more frames on the Display in the \textit{Abstract Light + Symbol} ($M=1.88, SD=2.21, p=.002$) and the \textit{Abstract Light + Text} ($M=1.93, SD=2.08, p=.001$) conditions than in the \textit{No Visual} condition ($M=0.90, SD=0.92$). We also observed an interaction effect of Group $\times$ Visual $\times$ Audio ($F(3,210) = 2.672, p = .048, \eta_{p}^{2}=.037$); however, post-hoc analysis did not yield any significant results. The duration participants spent on the Display can be found in \autoref{fig:BL_BD}b.

\begin{figure*}[t]
    \centering
    \begin{minipage}{0.49\textwidth}
        \centering
        \includegraphics[width=\linewidth]{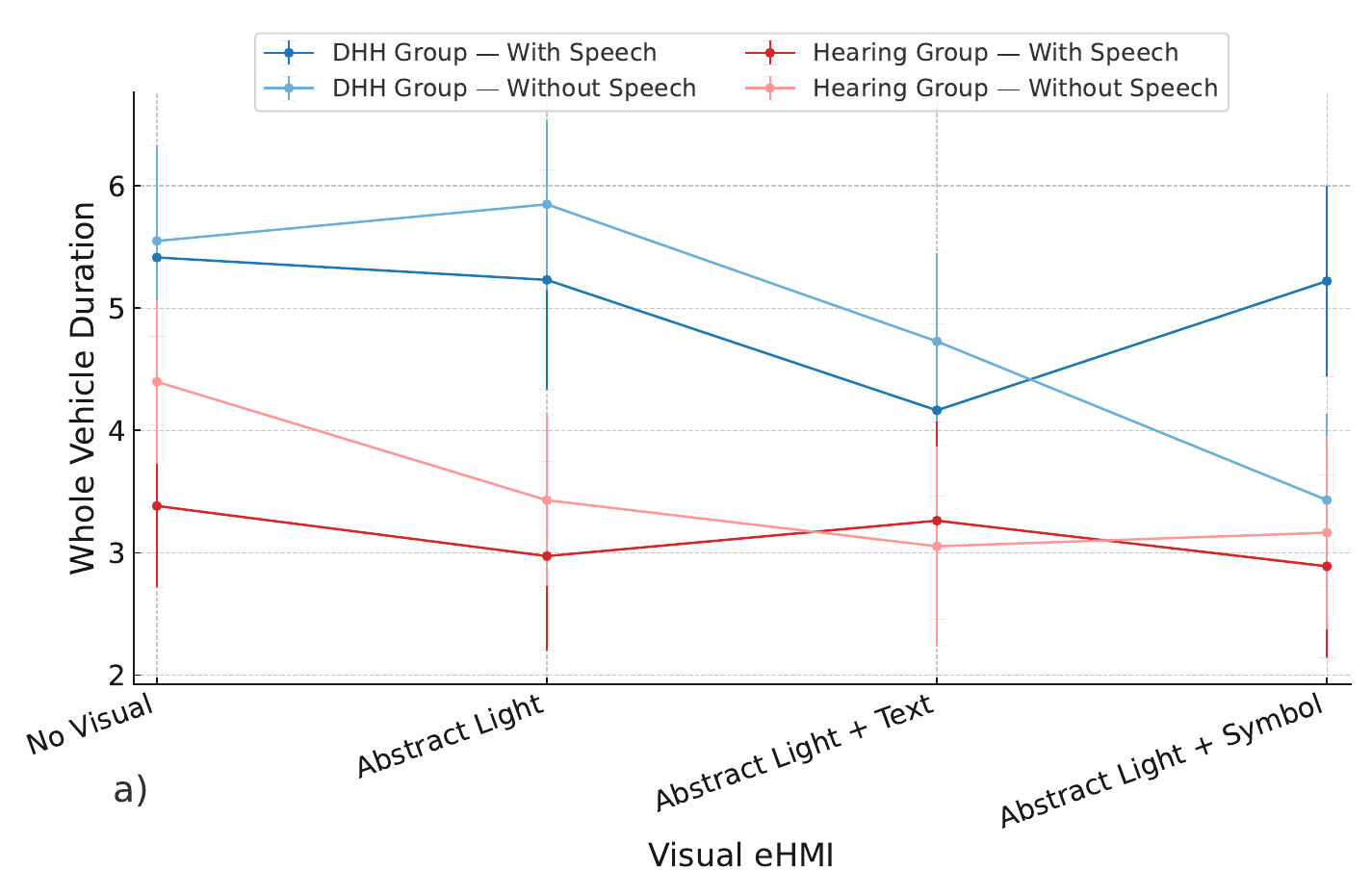}
    \end{minipage}
    \hfill
    \begin{minipage}{0.49\textwidth}
        \centering
        \includegraphics[width=\linewidth]{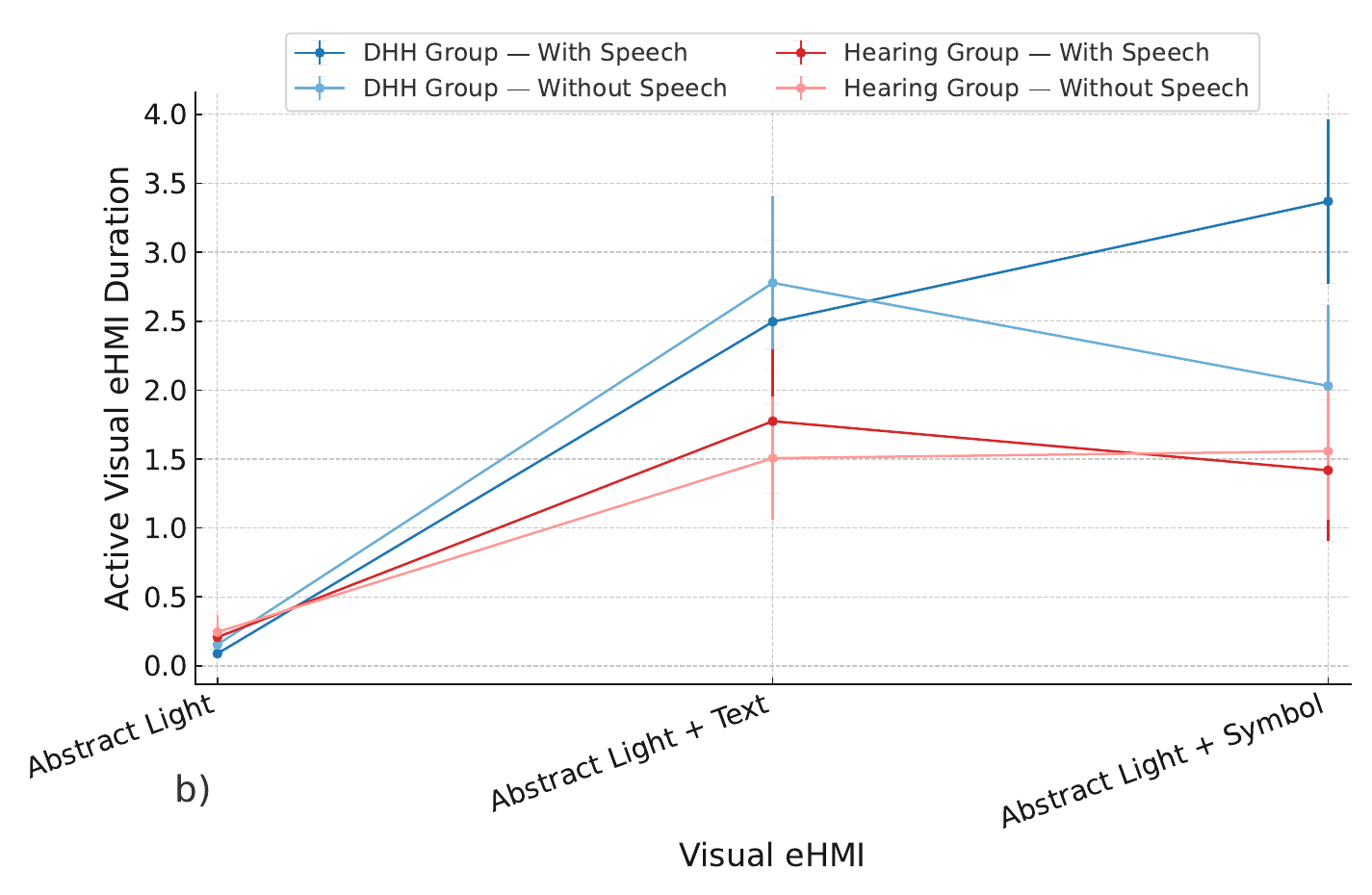}
    \end{minipage}
    \caption{a) Whole Vehicle Duration and b) Active Visual eHMI Duration by Visual × Audio eHMI conditions, separated for DHH and Hearing groups, with and without speech. Error bars show $\pm$SE.}
    \label{fig:wholeCar_active_ehmi}
    \Description{Data figure showing a) Whole Vehicle Duration and b) Active Visual eHMI Duration by Visual and Audio eHMI conditions for the DHH group.}
\end{figure*}

\textbf{Whole Vehicle Duration}. \autoref{fig:wholeCar_active_ehmi}a shows duration of participants in the Hearing and DHH groups spent on the whole vehicle. ANOVA tests yielded significant main effect of Visual ($F(3,210) = 5.330, p = .001, \eta_{p}^{2}=.071$) and Group ($F(1,30) = 4.271, p = .047, \eta_{p}^{2}=.125$). Post-hoc pairwise comparisons for the main effect Visual showed that participants looked significantly more time on the vehicle in the \textit{No Visual} condition ($M=4.69, SD=2.83$) than in the \textit{Abstract Light + Text} condition ($M=3.80, SD=3.06, p=.013$), the \textit{Abstract Light + Symbol} ($M=3.68, SD=3.09, p=.003$). Post-hoc pairwise comparisons for the main effect Group showed that participants in the DHH group ($M=4.95, SD=2.99$) looked at the vehicle significantly more time than participants in the Hearing group ($M=3.32, SD=2.94, p=.048$).

\textbf{Active Visual eHMI Duration}. We conducted further analysis to explore how long participants looked at active eHMI. ANOVA tests yielded significant main effect of Visual ($F(2,150) = 74.271, p < .001, \eta_{p}^{2}=.498$). Post-hoc pairwise comparison results suggest that participants spent more time on the active visual eHMI in \textit{Abstract Light + Text} and \textit{Abstract Light + Symbol} than \textit{Abstract Light} (both $p < .001$). We also observed interaction effect between Visual $\times$ Group ($F(2,150) = 8.097, p < .001, \eta_{p}^{2}=.097$), Visual $\times$ Audio ($F(2,150) = 3.305, p = .039, \eta_{p}^{2}=.042$), and Visual $\times$ Audio $\times$ Group ($F(2,150) = 4.550, p = .012, \eta_{p}^{2}=.057$). The post-hoc results of these three interaction effects confirmed the same post-hoc results of the Visual main effect. In addition, we also observed a significant interaction effect of Audio $\times$ Group ($F(1,150) = 4.485, p = .025, \eta_{p}^{2}=.029$), but no significant post-hoc results were found. The gaze duration spent on active eHMI component can be found in \autoref{fig:wholeCar_active_ehmi}b.

\begin{figure*}[t]
    \centering
    \begin{minipage}{0.49\textwidth}
        \centering
        \includegraphics[width=\linewidth]{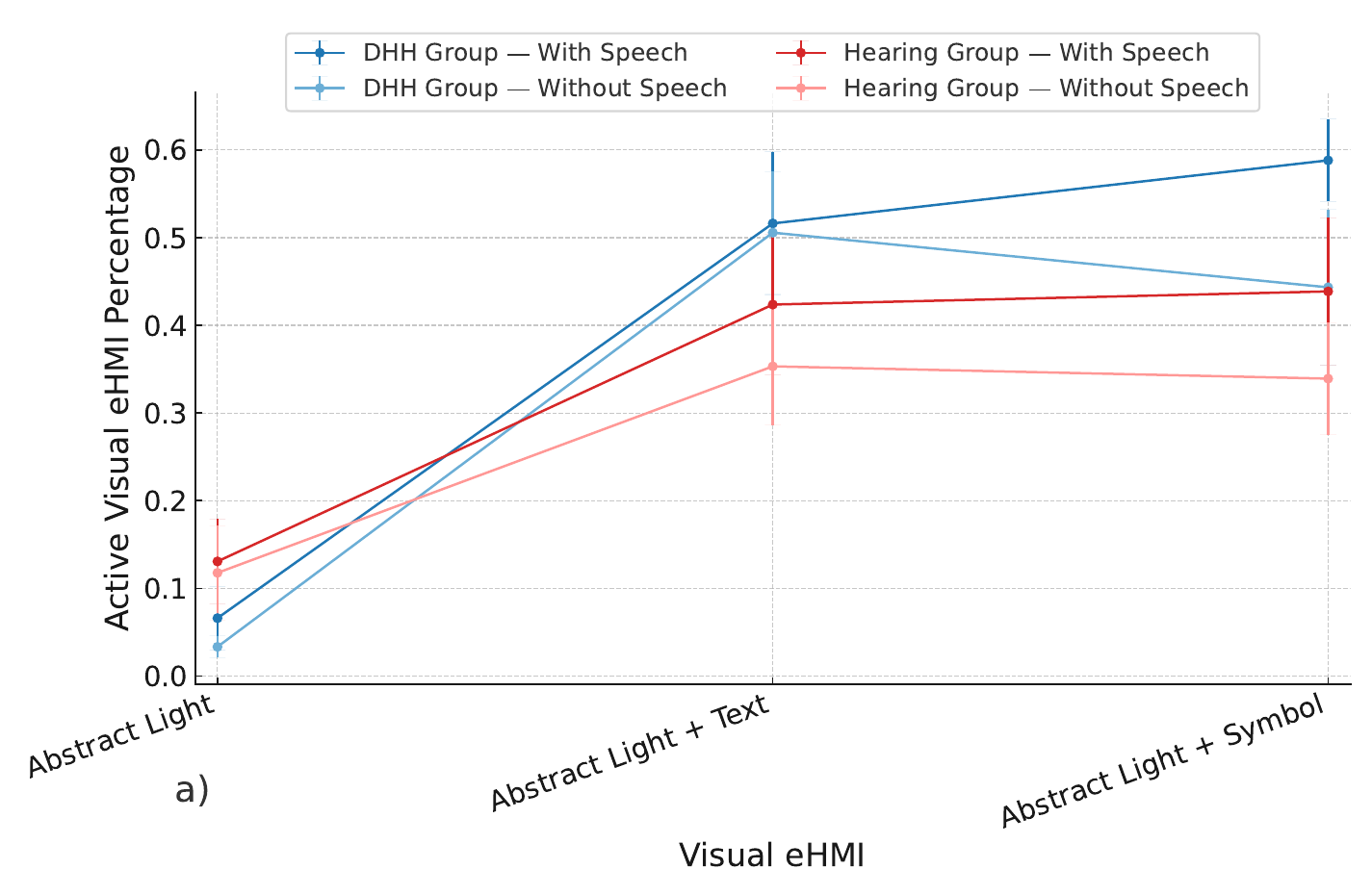}
    \end{minipage}
    \hfill
    \begin{minipage}{0.49\textwidth}
        \centering
        \includegraphics[width=\linewidth]{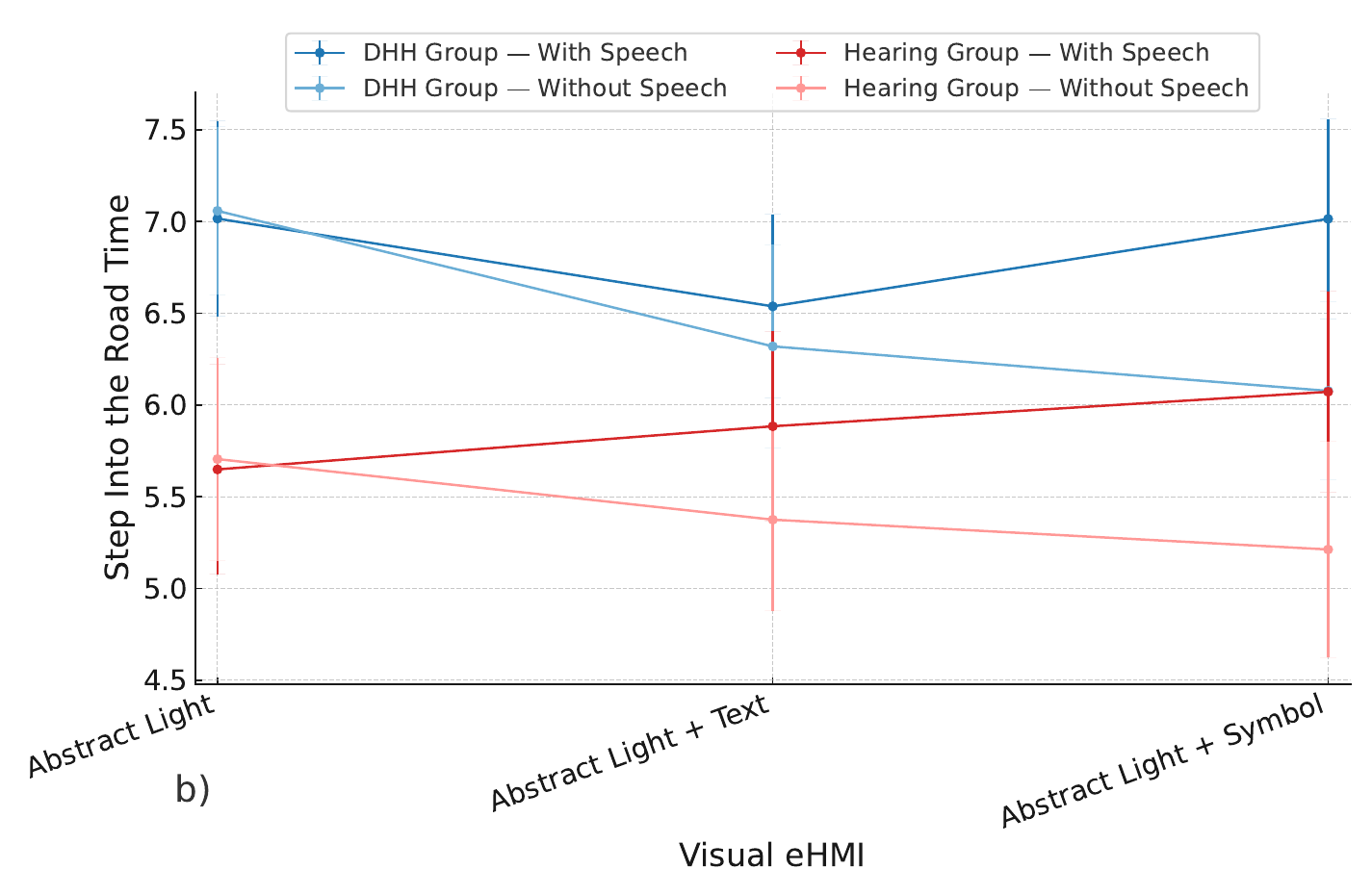}
    \end{minipage}
    \caption{a) Active Visual eHMI Percentage and b) Step Into the Road Time by Visual × Audio eHMI conditions, separated for DHH and Hearing groups, with and without speech. Error bars show $\pm$SE.}
    \label{fig:Percentage_step_in}
    \Description{Data figure showing a) Active Visual eHMI Percentage and b) Step Into the Road Time by Visual and Audio eHMI conditions for the DHH group.}
\end{figure*}

\textbf{Active Visual eHMI Percentage}. \autoref{fig:Percentage_step_in}a illustrates percentage of gaze duration on active eHMI component among the whole vehicle. 
ANOVA tests yielded significant main effect of Visual ($F(2,150) = 57.038, p < .001, \eta_{p}^{2}=.432$) and an interaction effect between Visual $\times$ Group ($F(2,150) = 6.710, p = .002, \eta_{p}^{2}=.082$). Both post-hoc pairwise comparisons showed the same results where participants spent a higher percentage of active visual eHMI components in \textit{Abstract Light + Text} and \textit{Abstract Light + Symbol} than in the \textit{Abstract Light} condition ($p<.001$). 

\subsubsection{Movement Data} We did not encounter a trial of participants crossing in front of the manual vehicles; meanwhile, no crashes were observed. Both data sets below covered 512 trials (32 participants $\times$ 8 conditions $\times$ 2 repetition) of crossings.

\textbf{Step Into the Road Time}. ANOVA tests yielded significant main effect of Visual ($F(3,210) = 5.684, p < .001, \eta_{p}^{2}=.075$). Post-hoc pairwise comparisons for the main effect Visual revealed that participants spent significantly longer time to step into the road in the \textit{No Visual} condition ($M=7.10 s, SD= 2.50 s$) than the \textit{Abstract Light} ($M= 6.36 s, SD= 2.18 s, p=.045$), the \textit{Abstract Light + Text} ($M= 6.03 s, SD=2.07 s, p=.002$), and the \textit{Abstract Light + Symbol} ($M=6.09 s, SD=2.21 s, p=.006$) conditions. The value for each condition across participants in the Hearing and DHH groups can be found in \autoref{fig:Percentage_step_in}b.

\textbf{Early Step Into the Road Count}. ANOVA tests yielded significant main effect of Visual ($F(3,210) = 2.699, p = .047, \eta_{p}^{2}=.037$). However, post-hoc pairwise comparisons showed no significant difference between the visual eHMI conditions, suggesting the number of times participants stepped onto the road before the vehicle fully stopped was similar across visual eHMI conditions.

\subsection{Necessity and Reasonability for Visual and Auditory eHMIs}
On average, participants in the Hearing group stated that there is a necessity for visual eHMIs ($M = 5.88, SD = 1.31$) and that their use is reasonable ($M = 5.69, SD = 1.49$). For auditory eHMIs, participants also perceived a moderate necessity ($M = 5.06, SD = 1.73$) and found them reasonable ($M = 5.31, SD = 1.49$). DHH Participants rated the necessity of visual eHMIs slightly higher ($M = 6.44, SD = 0.51$) and found them to be reasonable ($M = 6.31, SD = 0.70$). However, ratings for auditory eHMIs were lower, with perceived necessity ($M = 4.31, SD = 2.15$) and reasonableness ($M = 4.44, SD = 2.10$) showing more variability.

We were interested in understanding whether participants value the Visual and Auditory eHMI differently and whether hearing condition would impact this; therefore, we explored the impact of Modality (Visual vs Auditory) and Group (Hearing and DHH) on the necessity and reasonability ratings. As the data are not normally distributed, we employed ART ANOVA. ANOVA tests yielded significant main effect of Modality ($F(1,30) = 17.940, p < .001, \eta_{p}^{2}=.374$) on necessity ratings. Post-hoc analysis suggested that the necessity ratings for the visual eHMIs were significantly higher ($p<.001$) than the auditory eHMI. We could not find the main effect of the Group or any interaction effect. Regarding reasonability, ANOVA tests yielded a significant main effect of Modality ($F(1,30) = 9.48, p = .004, \eta_{p}^{2}=.240$). Post-hoc analysis suggested that the reasonability ratings for the visual eHMIs were significantly higher than the auditory eHMI ($p=.004$). We could not find the main effect of the Group or any interaction effect.

\subsection{Ranking}
The ranking of visual designs shows a preference for the \textit{Abstract Light + Symbol} ($M=1.66, SD=0.87$; 19 ranked it first while 8 ranked it second) and the \textit{Abstract Light + Text} ($M=1.66, SD=0.48$; 12 ranked it first while 18 ranked it second). They were followed by the \textit{Abstract Light} ($M=2.81, SD=0.74$; 1 ranked it first and 3 ranked it second, while dominantly ranked 3rd with 24 votes) and Base ($M=3.88, SD=0.34$; 28 ranked it last and never ranked 1st or 2nd). Friedman's Test showed that there was a statistically significant difference in the ranking data among the eHMI concepts ($\chi^2(5)=65.51, p <.001$). Post-hoc analysis showed that the \textit{Abstract Light + Symbol} was rated significantly better than the \textit{Abstract Light} ($p=.004$) and the \textit{No Visual} ($p<.001$). Similarly, the \textit{Abstract Light + Text} was rated significantly better than the \textit{Abstract Light} ($p<.001$) and Base ($p<.001$). In addition, we found that the \textit{Abstract Light} was rated significantly better than Base ($p<.001$).

Regarding the sound preference. 19 out of 32 participants ranked Speech as their first choice, while 13 ranked Without Speech first. The average rank score for Speech was 1.41 ($SD = 0.50$), compared to 1.59 ($SD = 0.50$) for Without Speech. The difference between the ratings was not significant. 

\subsection{Qualitative Results}
We conducted thematic analysis \cite{Braun01012006} with inductive coding on the interview transcripts (auto-transcribed by Teams and corrected by an author). In our reporting, we reported the insights related to each of the evaluated eHMIs, auditory eHMIs, and multi-modal eHMIs, with combined-visual eHMIs, along with selected participants' quotes (Hearing: P1-P16; DHH: P17-P32).

\subsubsection{Reflection on Decision Making strategy} Several participants (N=22; Hearing: 11, DHH: 11) mentioned that they either mainly rely on the visual eHMIs "\textit{Relied mainly on visual cues (faster)}" [P4], "\textit{Relied on visual cues; did not use audio cues}" [P17] or prioritized it over other cues "\textit{relied more on visual cues than audio}" [P22] which could be due to their disability "I\textit{ don't trust my hearing}" [P18], "\textit{primarily visual eHMI; sound was secondary help if attention was low}" [P12]. 3 participants (Hearing: 2, DHH:1) mentioned that they mainly relied on implicit communication (i.e., "\textit{the car slows down}" [P7, P10, P26]. However, they also acknowledged that certain eHMI design (i.e., "\textit{the walking man symbol helped me and made my decision}" [P26]. 1 participants (DHH:1) mentioned that they did not have a priority among these eHMIs, either eHMI work for them, they wait "\textit{any signal from the vehicle, visual or audio cue}" [P19] and 1 participant (Hearing: 1) mentioned they would rely on both visual and auditory eHMI. 2 people from DHH group mentioned that due to the nature of deafness, they would rather check everything before they made the crossing decision (implicit cues, visual, auditory eHMIs, potential upcoming vehicle from other side of the road) [P32]. 2 participants (Hearing: 1; DHH:1) mentioned regardless of everything, they would wait till the vehicle fully stop. 

\subsubsection{Reflection on Visual eHMIs}
\textbf{Abstract Light}: 
Several participants (N=25; Hearing: 9, DHH: 16) expressed positive feeling of the light component and its design, suggesting it "\textit{can see it from a distance}" [P3, P10, P16], "\textit{aesthetically pleasing}", and "\textit{reassuring that I'm seeing you}" [P4, P29, P31]. As for the design pattern, participants expressed they "\textit{liked the blinking effects}" [P2, P8, P10, P14, P23, P30], it felt like the lights "\textit{giving me the indication that it's gonna decelerating and it's gonna stop}" [P4, P14, P26]. 

However, several participants had raised concerns about the light component and its design, mainly because they believed it could be confusing, particularly when they first perceived the light design alone. A few participants (N=8; Hearing: 3, DHH: 5) stated that the light "\textit{was not understandable, not sure what the light was telling me to do}" [P5, P22, P31], and some participants (N=4; Hearing: 2, DHH: 2) admitted that they "\textit{had to guess the meaning of the light}" [P6, P12, P21, P25]. Additionally, a few participants (N=8; Hearing: 3, DHH: 5) expressed concerns that "\textit{it may not be clear for everybody}" [P16], "\textit{some people may misinterpret the flashing}" [P26], "\textit{could confuse people, is it part (design) of the car? Is it aesthetics?}" [P32], "\textit{might be difficult for people who don't know what the light (is) for}" [P3, P9, P22, P23, P27].

We observed a strong preference for Combined-Visual or Multi-Modal Cues (i.e., combining visual, textual or symbolic and even auditory cues), 8 participants explicitly mentioned it (Hearing: 2, DHH: 6). P15 said, "\textit{It's more visual to have the light, but if you have something written or the symbol, it's even better}". A primary reason of the claim is due to the fact of confusion (there will be previously discussion on confusion) made by lights own it's own, "\textit{lights on its own ... it's telling me what the car is doing, but it's not communicating to to me what it wants me to do.}" [P28].

\textbf{Abstract Light + Text}: 
Majority of participants (N=30; Hearing: 15, DHH:15) expressed positive feeling of this combination and their design, stating it was "\textit{clear}" [P2, P4, P7, P10, P22, P25, P30], "\textit{easy to understand}" [P2, P18, P30], "reassuring and confidence-building" [P4, P10, P20, P23, P31], and "\textit{reliable and helpful}" [P5, P17, P23, P26, P29]. Despite the majority of the people saying this combination was great, 3/32 participants expressed that using the Abstract Light itself is enough (P1, P2, P10) and "\textit{you always see the light first}" [P2], with 1 participant arguing that "\textit{I do not think the light was necessary}". 8/32 participants raised concerns towards the difficulties with the use of Text, as (1) it generates language barrier for "\textit{some Deaf sign language people, English is not a preferred language for them}" [P26], (2) seeing the Text from distance could be challenge, with P31 stated "\textit{not necessarily be able to read it at a distance}", and (3) it could be "\textit{too much information (to see)}" [P1] and "\textit{takes a bit more time (to process), it slows the reaction time}" [P23],--however, they later admitted that "\textit{it is for safety}" to keep the Text. 

\textbf{Abstract Light + Symbol}: 
Twenty-four participants (Hearing: 10, DHH: 14) expressed positive feelings towards this condition and its design. They stated that this was "\textit{Clear}" [P23], "\textit{probably less time to process than text}" [P27], "\textit{reassuring}" [P4, P20]. Many people believe the benefits are due to the signs used looking like "\textit{traffic signs}" (N=7). In addition, they believed that using signs could work for people who cannot read [P2, P5, P9, P20], Children [P26], or non-native speakers [P1, P3]. However, 19 participants raised negative comments for Abstract Light + Symbol, while fewer participants (N=8) raised negative comments for Abstract Light + Text. For instance, Almost half of the participants (N=14) expressed that the Symbol could be unclear or contains abstract meaning (especially the "\textit{standing man}"), they said there was an initial learning curve "\textit{the first time, I wasn't quite sure what it meant}" [P19, P31], people could misinterpretation the Symbol or had wrong assumptions (N=5). In addition, a few participants (N=7) expressed that Symbol is harder to understand than Text as "\textit{Text is more self-explanatory}" [P10, P11], while Symbol "\textit{you have to actually interpret it; harder at a glance}" [P14].

\subsubsection{Reflection on Auditory eHMIs}
All Hearing group participants (N=16) could hear the auditory eHMI from the vehicle. However, this was not the case for DHH participants; only 9/16 stated they could hear the auditory eHMI. 3 HoH participants struggled to hear the auditory eHMI, with [P17] stating that he could hear "Stopped" but not the "Deceleration", where [P18, P32] both admitted that "\textit{I'm struggling}", and they focused more on the visual cues as "\textit{visual is much more effective}" [P32] and are "\textit{used to using visuals}" [P18]. All Deaf (BSL users) (N=4) were unable to hear the words, two of them were unable to hear anything (including 1 wearing hearing aids); the other two were able to pick up a sound, but unable to understand the meaning of the sound---"\textit{I heard a little bit, but I don't know clearly what they are saying, there is a sound, but not a word and what the word is}" [P25] with [P29] sharing the similar quote "\textit{I didn't understand that, I didn't know what was saying}". However, both also noted that they were able to interpret the noise in the Abstract Light + Text condition "\textit{I could hear that it said stopped and that helped me go.}" 

Over half of the participant sample (22/36) mentioned that the auditory eHMI brought benefits to them when they interacted with the AV, "\textit{when audio is provided it felt good that the system tells people what the car is doing}" [P4, P21], "\textit{audio served as a confirmation}" [P14, P21, P27], "\textit{help awareness and decision-making}" [P12, P19], "\textit{reinforced the (visual) message}" [P3, P8, P31], "\textit{increase confidence}" [P13, P22, P31], and they just "\textit{like/love it}" [P5, P10, P11]. However, 1 DHH participant [P17] disliked it "\textit{hindrance rather than a help}" because "\textit{Sound can interrupt or break up other processes, making it harder to concentrate}" and "I only captured the word stop" due to my hearing loss. And 1 Hearing participant disliked it as "\textit{It feels uneasy and overloading}" [P14].

In addition to these benefits to themselves, 12 participants mentioned that having auditory eHMI could be helpful for "\textit{people who are blind/cannot see/vision loss}" [P2, P15, P16, P26, P28, P29], "\textit{people who have bad eye sight}" [P1, P7], "\textit{children}" [P5], "\textit{colour blind}" [P17, "\textit{people who don't look up}" [P2], and the distracted people such as "\textit{people wearing headphones}" [P3] or "\textit{everyone if they are not paying attention}" [P9].

\subsubsection{Reflection on Design Requirements}
\textbf{State-Transition}: All participants noticed the changes in the design in different states (i.e., slowing down and fully stopped). Nineteen participants (Hearing: 8; DHH: 11) consistently valued state-transition feedback in eHMI design from the slowdown to fully stopped, noting it improved clarity, trust, and decision-making. Text and light cues that explicitly reflected vehicle state were described as intuitive and reassuring [P3, P6, P12, P16]. The progression helped users anticipate vehicle actions, especially when matched with natural behaviours. Participants also recommended (1) simplifying phrasing (e.g., avoiding technical terms like “\textit{deaccelerating}”) to ensure quick comprehension, and (2) keeping the message of each state-transition consistent across different modalities.

\textbf{Multi-Modal eHMI}: We did not ask questions directly related to multi-modal eHMIs; however, 12 participants (Hearing: 5; DHH: 7) explicitly reported the need for multi-modal eHMIs (i.e., using visual and auditory together), with 1 participant also explicitly suggesting to include haptic feedback for the users through "\textit{a bracelet or an IoT like the Apple Watch that will vibrate}" [P30]. They believe multi-modal eHMI increases clarity of communication, improves reaction times, and ensures inclusivity for people with diverse abilities and needs. For instance, P22 stated that "\textit{everything (seeing and hearing the vehicle slow down) matched could increase confidence and sense of trust}", while P13 said "\textit{it gave double confirmations, increasing confidence to cross}". In complex settings or to someone who might not be able to receive a cue from the vehicle, multi-modal eHMI increased accessibility through allowing people to choose their reliable source of information, which was confirmed by our participants (N=6; DHH: 6) with P18 said "\textit{My hearing is quite bad. So I rely on visuals more than anything else}" and P29 explained that "\textit{I relied more on the visual. For some people, maybe rely on the audio, people who are blind and so on, but for me, as a deaf person, no}." Despite there are benefits of having sound stimuli, 8 participants (Hearing: 1, DHH:7) raised concerns on how effective this could be in the real world, "\textit{in real world with more noise, more information, more people}" [P4], "\textit{busy traffic or noisy environment, the audio might not be audible}" [P20], could be "\textit{hard to detect}" [P25] as "\textit{background noise could block the sound}" [P29, P32], "\textit{making them unhelpful}" [P29] and "\textit{not reliable}" [P32]. In addition, 3 participants argued that having multiple vehicles speak could exacerbate noise pollution and be perceived as "\textit{annoying}" [P1], "\textit{overwhelming}" [P7, P28].

\textbf{Combined-Visual eHMI}: Seven participants (Hearing: 1; DHH: 6) explicitly mentioned that there should be multiple visual cues on the visual eHMI design than simply using an Abstract Light eHMI, combining Abstract Light with Text or Symbol displays, to offer clearer, more informative, and more reassuring communication. They help communicate the vehicle’s intention, increase user confidence over time, and align with familiar public signage conventions, making them more intuitive for diverse pedestrian groups.

\section{Discussion}
eHMI research still faces a significant lack of focus on evaluating these concepts with disabled people, especially DHH people. Our study is the first VR study that evaluates the effects of visual and auditory eHMI on DHH people's experiences (trust, acceptance, perceived safety, mental load) and behaviour (gaze behaviour, step-in road time, early step into the road count) while comparing it to hearing people. Our RQs were: 

\textbf{RQ1: To what extent do the ratings for experience and behaviour differ between participants in the Hearing group and in the DHH group?} We found no significant difference between experience-related ratings from the Hearing and DHH group participants. However, crossing behaviour was affected when crossing in front of the AV in the tested scenario. Participants in the DHH group spent significantly more time looking at the AV than the Hearing group. \citet{street_crossing_disability} had found that DHH pedestrians often had to exercise extreme caution towards manual drivers because of their inability to perceive auditory stimuli. Due to the absence of human drivers in AVs, we suspect DHH spent more time looking at the AV due to caution. This was partly supported by our qualitative data, where 2 participants mentioned they would wait till the AV fully stops, and 2 participants mentioned they would check everything before making the crossing decision, when nobody in the Hearing group mentioned similar comments. 

We did notice interesting group-specific patterns. DHH participants rated \textit{Abstract Light} significantly higher in usefulness than the No Visual condition, whereas this difference was not observed in the Hearing group. Conversely, Hearing group participants rated \textit{Abstract Light} significantly higher than No Visual for perceived safety, but this pattern did not hold for DHH participants. These nuanced interaction patterns suggest that while the two groups provide similar overall experience ratings, they differ in how specific visual eHMIs support their crossing experience.

\textbf{RQ2: What impact do the visual eHMIs have on pedestrians regarding experience and behaviour?} In line with early works in video-based research~\citep{10.1145/3473856.3474004,multi_modal_dey}, VR research~\citep{scalability_Mark,Hollander_ehmi}, and real world Wizard-of-Oz research~\citep{FAAS2020171,matthews2017field}, our results show that providing visual eHMI (i.e., \textit{Abstract Light + Symbol} and \textit{Abstract Light + Text}) on the AV improves subjective crossing assessments (trust, usefulness, satisfying, and perceived safety). When compared to not having any eHMI, providing \textit{Abstract Light} alone (1) improved usefulness rating among DHH people, (2) improved usefulness and satisfying ratings when speech was not provided, and (3) improved perceived safety among hearing people but not for DHH people. 

Our gaze behaviour data further indicate that combined-visual eHMI positively impacted the participants' gaze behaviour; they spent more time on the active visual eHMI and less time on the AV. The crossing behaviour data suggest that participants spent less time stepping on the road, regardless of which visual eHMI was provided. Participants did not encounter any accidents. Therefore, our research clearly supports previous research in that having visual eHMIs assists in crossing decision making~\citep{Hollander_ehmi}. 

We observed a similar trend where the use of eHMI seems able to rivet and centralise pedestrians' visual attention to the active visual eHMI area, supporting previously reported findings in a video-based study~\citep{su14095633}. In addition, we noticed that when \textit{Text} and \textit{Symbol} were active, more visual attention was paid to these visual designs than to the \textit{Abstract Light}. One explanation could be that \textit{Text} and \textit{Symbol} were reported to be clearer and easier to understand when compared to \textit{Abstract Light}, according to the qualitative insights provided by the participants. Participants still paid some attention to the active \textit{Abstract Light} eHMI under \textit{Abstract Light + Text} and \textit{Abstract Light + Symbol} conditions, but they were much less than in the \textit{Abstract Light} condition alone. Under \textit{Abstract Light + Text} and \textit{Abstract Light + Symbol} conditions, participants commented that they check the \textit{Abstract Light} component first when the vehicle is far, which was supported by validating the recorded video with the participants' gaze movements---participants tend to focus on the \textit{Abstract Light} at a far distance and then focus more on \textit{Text} or \textit{Symbol}.

\textbf{RQ3: What impact does providing auditory speech-based eHMIs have on pedestrians regarding experience and behaviour?} For our tested scenario, providing Speech-based eHMI generally improves the crossing experiences (trust, usefulness, perceived safety) compared to not providing it. This finding supports the auditory eHMI literature conducted in video-based research~\citep{multi_modal_dey}, simulation research~\citep{Mark_Vision}, and real world Wizard-of-Oz research~\citep{BINDSCHADEL202359}. However, we did not observe a significant impact of providing Speech-based eHMI on the pedestrians' crossing behaviour, such as eye gaze behaviour and step into the road decision making, which is not in line with the real world Wizard-of-Oz research~\citep{BINDSCHADEL202359}, where participants made faster crossing decisions when the intention of the vehicle was played.

To answer \textbf{RQ3}, we confirm that, for the scenario tested in this study, providing Speech-based auditory eHMI positively impacts crossing experience but does not impact crossing behaviour. The crossing behaviour, especially the speed-related movement measurements, could be improved if an advice/instruction message was used rather than the chosen state/intention message~\citep{EISMA_text}; however, such message design could bring in liability and legibility issues~\citep{status_zhang,FAAS2020171} as discussed in the Section \ref{designRequirement}. 

It is also important to acknowledge that Speech-based auditory eHMIs may not be effective for all pedestrians, particularly sign language users whose first language is not English (i.e., the chosen language for this study) or those who cannot hear anything. Among the four sign language users we recruited, two were able to perceive the speech sounds. However, they could not interpret the message correctly in most conditions, except in the \textit{Abstract Light + Text} condition, where they could match the sound with the text, even though the spoken stopping message differed from the visual cue. For the other two participants who reported being unable to hear the speech, we conducted an additional check by replaying the sound through a headset after the experiment. Interestingly, one of these participants was able to hear the message when not engaged in the crossing task. We suspect this may be due to inattentional deafness~\citep{Macdonald2011}, a phenomenon where people fail to notice auditory stimuli (i.e., speech message) when concentrating on a demanding visual task (i.e., crossing).

Nevertheless, three out of four sign-language users believed that Speech-based auditory eHMI is still helpful and would benefit more people. The other sign language user said that he did not care too much, as he could not hear the sound at all. Future work could explore other forms of multi-modal eHMI (e.g., haptics).

\subsection{Decision Strategy}
Twenty-four participants reported relying primarily on explicit communication cues (Visual: 22; Either Visual or Auditory: 1; Both: 1) when making crossing decisions in front of the AV, whereas only 3 participants stated that they depended solely on implicit cues such as vehicle kinematics. This finding supports prior works~\citep{9575246,Tram_CHI_24,BINDSCHADEL202359} that the presence of eHMI improves subjective feelings and plays an important role in decision-making. This contrasts with research on understanding pedestrian-manual driven vehicles communication, where \citet{10.1145/3122986.3123009} found that explicit communication is rare to non-existent and does not play a significant role, implicit cues---particularly vehicle kinematics---played a more dominant role in making crossing decisions~\citep{10.1145/3239060.3239082,SUCHA201741}. We argue that even though pedestrians can cross only relying on implicit communication such as the AV's motions~\citep{10.1145/3342197.3345320}, the presence of active eHMIs is needed as it could attract pedestrians' attention and support better decision-making and a better crossing experience, especially during the slowdown process.

\citet{DEB2018135} suggested that visual eHMIs had more impact on the willingness to initiate crossing than auditory eHMIs. This is supported by our results, 22 out of 24 participants explicitly mentioned they relied on eHMI, indicating that they relied more on the visual than the auditory eHMI, while only 1 indicated the opposite. In addition, results of the necessity and reasonableness ratings also indicated that participants believed visual eHMIs were much more necessary and reasonable than the auditory counterpart. 

\subsection{Ecological Validity of Study Approach}\label{ecological validity}
Our design choices reflect a deliberate balance between experimental control and ecological relevance, which is consistent with prior eHMI research through VR (e.g., \citep{10.1145/3491102.3517571,Deb_stop,Hollander_ehmi}) and CAVE simulators \citep{LEE2022270}. 

As lack of realism (appearance, content, task, and setting) in virtual world could negatively impact the validity \citep{10.1145/1979742.1979621}, we ensured our study had a good realism by (1) using visually realistic vehicles and urban surroundings, (2) employing a familiar task (street crossing), and (3) situating the interaction in a typical urban road layout. In addition, to increase internal validity, we intentionally constrained elements that could confound participants’ responses: all vehicles had identical size/colour, yielding behaviours that were consistent, and no additional occlusion-inducing objects were present in the scene (e.g., parked vehicles, trees, bins). We also implemented two vehicles in the second lane at the beginning of the experiment to remind participants to check the left (most participants did check the left while crossing). We told the participants that both manual and automated vehicles may stop for them, although only the AV would stop for them. These implementations enable a clear interpretation of our study; they also come with the drawback of reduced ecological validity. Real world settings can often be more hazardous due to occlusions, variable approach vectors, multiple AVs, and mixed compliance ~\citep{wearther_lighting, crossing_visual}; the crossing logic may be more dynamic, depending on the weather and time of day. Therefore, although our research provides meaningful early-stage research with DHH people, further testing is needed to explore crossing in more complex road setting.

Despite these simplifications, our findings remain meaningful and transferable. As noted by \citet{Recarte2005ArrivalTime}, time-to-arrival estimates are consistent between video simulations and real-life situations. Early browser-based experiments reported that walking behaviour aligned closely with real world data \citep{10.1145/3334480.3382797}. \citet{SHEN2023101716} created a video-based tool to assess the safety of young pedestrians’ street-crossing behaviour and found these video-based assessments to be both valid and reliable. \citet{Fuest_vr_comparison} compared video-based, VR-based, and Wizard-of-Oz vehicles' approach and observed only minor descriptive differences between them.

\subsection{Practical Implications}
We revisit the high-level design requirements identified in our formative study, reflect on these in light of our observations, and propose the following practical implications:

\subsubsection{Include Various Populations in the Design and Evaluation} Despite most quantitative results being rated similarly (no significant difference) between participants in the Hearing and DHH groups, we found that gaze behaviour, like time spent on the vehicles, differed. In addition, many more DHH people asked for the need to have combined-visual eHMIs than participants in the hearing group. In line with~\citet{Mark_Vision}, we call for future works to include various populations in the design and evaluation of eHMIs to make an inclusive eHMI for all people.

\subsubsection{Display Key State and State-Transition of the Vehicle}
Our qualitative findings suggest that conveying a vehicle’s transition between states can enhance DHH pedestrians’ clarity, trust, and decision-making when anticipating vehicle actions. We therefore recommend that eHMIs explicitly indicate the state transitions (i.e., slowing down) and the stop in scenarios where vehicles yield to pedestrians. Consideration should be taken when applying (1) \textit{Text} for the visual design, technical terms such as decelerating or deaccelerating should be avoided in favour of plain, widely understood terms like slow or slow down; (2) Speech together with the \textit{Text} as the visual eHMI, the terminology should remain consistent across the visual and auditory channels.

\subsubsection{Avoid Traffic Style Symbols}
Traffic signs convey meaning by a combination of symbols, shapes (e.g., circles--give orders, triangles--warning, rectangles--provide information), and colours (e.g., Red--prohibition or danger, Blue--instructions or routes for specific traffic, Green--direction signs on primary routes). By avoiding input shapes and colours, we assumed the "Standing Man" symbol and the "Walking Man" symbol from typical traffic signs would not confuse participants that the symbols they see are not traffic signs. However, 7 out of 32 participants explicitly mentioned that they treated this as traffic-like signs, as we expected, but took "advice" of the "Walking Man" symbol. Although these symbols are easier for people, this may against our intention of not providing crossing advice, which could lead to liability and legibility issues \citep{status_zhang,FAAS2020171,iso23049}. Therefore, we encourage future work and design efforts to take this consideration (i.e., avoid using traffic style symbols) into account unless (1) broad public awareness or training campaign are in place to familiarise people with eHMIs~\citep{Tram_CHI_24}, or (2) regulations evolve to include AVs as part of future traffic infrastructure~\citep{mark_parked}.

\subsubsection{Enabling Multi-Modal eHMI}
Data from crossing experience, crossing behaviour, interview, and ranking showed a consistent preference for having a visual eHMI, suggesting all multi-modal systems should rely on visual as the base. Although the ranking data does not show a clear support for the claim that participants prefer to have Speech-based auditory eHMI over not having it, there is a general trend that employing Speech-based auditory eHMI enhances their crossing experiences (i.e., trust, acceptance-usefulness, perceived safety). Therefore, our data support the notion that eHMI should be multi-modal and are suitable for people with DHH.

The benefits of having multi-modal eHMI were positively discussed by our participants and related works~\citep{multi_modal_dey,Mark_Vision}. However, transport noise ranks among Europe's top three environmental health threats; over 20$\%$ of Europeans are exposed to harmful transport noise levels~\citep{EEA2025Noise}. Considerations are needed when employing auditory eHMI as part of the multi-modal eHMI; designers should ensure auditory eHMI comply with WHO's environmental noise guidelines~\citep{WHO2018NoiseGuidelines} and are noticeable under different background noise levels~\citep{10.1145/3409120.3410646}, as there could be an impact of background noise on auditory eHMI \citep{Xu_CHI_26_noise}.

\subsubsection{Enabling Combined-Visual eHMIs}
Literature and our qualitative feedback suggest that singular visual communication designs have limitations. Using \textit{Abstract Light} alone to express intention may be unintuitive~\citep{info13090420}. At the same time, \textit{Text} appeared to be more easily understood than light~\citep{EISMA_text}, it may require more visual attention~\citep{BAZILINSKYY2021103450,info13090420} and is less noticeable than \textit{Abstract Light} in far. \textit{Symbols} have fewer language barriers and are easier to observe than \textit{Text}~\citep{distance_read}, but are less noticeable at a distance than \textit{Abstract Light}. Most critically, pedestrians may treat them as traffic signs. 

Based on our findings with DHH people, we suggest future eHMI research/design to enable combined-visual communications as they lead to combined benefits, compensate for the limitation of singular visual communication, and outperform singular visual communication designs. Considering Symbols used in our study may be treated as traffic signs when participants are not educated with the meaning of eHMI, leading to liability issues \citet{status_zhang}. We recommend implementing \textit{Abstract Light + Text} as a starting point. However, this recommendations need to be further tested with people with other disabilities. Additionally, we did not test other visual methods, such as \textit{Situational Awareness} and \textit{Road-based Projection}, through VR simulation; further work should explore the combined-visual eHMI ideas with these designs.

\subsection{Limitations and Future Work}
This research has some limitations, which also highlight promising directions for future work. Given the limited number of prior research involving DHH participants, we followed previous study~\citep{DEB2018135} and adopted a single controlled scenario featuring a non-signalised crossing. This setup effectively minimised external distractions, allowing participants to focus on the eHMI features being studied. However, such a controlled environment may constrain ecological validity (see Section \ref{ecological validity}). Future research should investigate more complex contexts, such as (1) mixed-traffic with more AVs in the scene, (2) crossings shared with other pedestrians~\citep{scalability_Mark}, and (3) crossing in intersections or zebra crossings, to examine how these factors influence pedestrians’ crossing experience and behaviour. 

We only used the \textit{Abstract Light} as the singular visual eHMI concept---the best singular eHMI candidate found through our formative study. We did not test \textit{Text} or \textit{Symbol} independently as singular eHMI. In addition, the lower ranked visual concepts from formative study (i.e., \textit{Situational Awareness}, \textit{Road-based Projection}, and \textit{Anthropomorphic}) were not evaluated in VR as having them would significantly increase the duration of our VR study, hence not feasible to our participants, we encourage future work to further explore them in VR and field studies. For auditory eHMI, we only evaluated \textit{Speech}, as it was considered more inclusive for low vision or blind people~\citep{Mark_Vision} and was preferred over other audible cues in prior research~\citep{DEB2018135,Hudson_verbal}. However, \textit{Speech} may not be an ideal medium for all DHH people, especially those whose primary language is sign language. Future work could benefit from deeper collaboration with DHH advocacy groups and from recruiting a more diverse pool of DHH participants—representing different linguistic and cultural backgrounds—to ensure the findings resonate globally. Moreover, exploring other modalities such as non-verbal auditory cues (e.g., music or bells \citep{multi_modal_dey,Xu_CHI_26_noise}) or haptic feedback could further enhance inclusivity and accessibility.

Our study involved only a single experimental session and was only tested in the UK. Future research could adopt a longitudinal design, extend participation to other groups of disabled pedestrians~\citep{Mark_Vision,cognitive_impairment,Asha_wheelchair}, and test the implications in different countries and cultures to determine if they would apply in the long term, across various countries and cultures, and for other types of disabilities.

\section{Conclusion}
In this research, we first conduct a formative study via focus groups with DHH people and relevant stakeholders to (1) inform the selection of visual eHMI candidates and (2) identify foundational design needs. Then, we investigate the effect of Visual eHMI (\textit{No Visual}, \textit{Abstract Light}, \textit{Abstract Light + Text}, \textit{Abstract Light + Symbol}) and Speech-based Auditory eHMI (\textit{With Speech}, \textit{Without Speech}) for AV-pedestrian communications in a VR study, with additional interest to explore differences regarding crossing experience and behaviours among participants in the hearing and DHH groups. Our results suggest three conclusions for the crossing scenario we evaluated: (1) Considerations should be paid to DHH people regarding eHMI design, as they show different crossing behaviour (i.e., gaze duration on the AV). (2) Visual eHMIs, singular or combined-visual, improve both crossing experiences and crossing behaviour. (3) Auditory eHMI, like Speech-based auditory cues used in our experiment, might not help with crossing behaviour, but improve the overall crossing experiences. We also proposed five valuable implications that can enhance the inclusivity of AV-pedestrian communications accessible to DHH people.

\begin{acks}
The authors thank all participants for their time. This work was funded by the Royal Society (RG\textbackslash R1\textbackslash 241114).
\end{acks}

\bibliographystyle{ACM-Reference-Format}
\bibliography{sample-base}

\end{document}